\documentclass[aps,prl,twocolumn,amsmath,amssymb,superscriptaddress,hidelinks]{revtex4-1}

\usepackage{bm}
\usepackage{verbatim} 

\usepackage{graphicx}
\usepackage[usenames,dvipsnames]{color}
\usepackage[colorlinks]{hyperref}
\hypersetup{
  colorlinks,
  citecolor=blue,
  linkcolor=blue,
  urlcolor=blue}

\newcommand{\be}{\begin{equation}}
\newcommand{\ee}{\end{equation}}
\newcommand{\bea}{\begin{eqnarray}}
\newcommand{\eea}{\end{eqnarray}}

\renewcommand{\vec}[1]{{\bf #1}}

\begin{document}

\renewcommand{\hbar}{\mathchar'26\mkern-9mu h}

\title{Skew-scattering Pockels effect and metallic electro-optics in gapped bilayer graphene}

\author{Da Ma} 
\affiliation{Division of Physics and Applied Physics, School of Physical and Mathematical Sciences, Nanyang Technological University, Singapore 637371}
\author{Ying Xiong} 
\affiliation{Division of Physics and Applied Physics, School of Physical and Mathematical Sciences, Nanyang Technological University, Singapore 637371}
\author{Justin C.W. Song}
\email{justinsong@ntu.edu.sg}
\affiliation{Division of Physics and Applied Physics, School of Physical and Mathematical Sciences, Nanyang Technological University, Singapore 637371}

\begin{abstract}

We argue that a range of strong metallic electro-optic (EO) effects can be naturally realized from non-Drude dynamics of free carriers in metals. In particular, in clean metals we identify skew-scattering and a ``Snap'' (third-order derivative of velocity) dominating the Pockels and Kerr EO behavior of metals in the clean limit. Strikingly, we find that both Pockels and Kerr EO in metals play critical roles in metallic EO phenomena: for instance, metallic Pockels and Kerr EO effectively compete to produce a field-activated birefringence that is non-reciprocal in applied DC fields. Similarly, both contribute to sizeable field-induced modulations to transmission and reflection across a range of frequencies. We find metallic EO effects can be naturally realized in layered 2D materials such as gapped bilayer graphene producing pronounced values of EO coefficients in the terahertz -- an interesting new metallic platform for terahertz electro-optic modulation.  
\end{abstract}

\maketitle

Electro-optic (EO) effects, where a material's linear optical properties are manipulated by an applied DC electric field, play a critical role in modern optical device components~\cite{Shen,Jenkins,New,Vorobev1979,Shalygin2022}. Amongst the most prominent EO effects are the Pockels and nonlinear Kerr effects that alter the dynamical conductivity (or equivalently, dielectric response) of a material at linear or second-order in applied DC electric field respectively. These EO effects enable to modulate the amplitude and polarization of light, deflect its direction of propagation, and can be used for sensors of intensity and phase, and even for amplification of light~\cite{Shi2023,Rappoport2023,Hakimi2023,Morgado2024}. 

While traditional EO effects focus on interband processes~\cite{Sipe2000,Vasko2006,Strikha2010,Margulis2017,Sabbaghi2018}, recent attention has turned to exploiting nonlinearities in noncentrosymmetric metals at lower frequencies (e.g., terahertz). Unlike insulators, free carriers in metals make intraband metallic EO possible. For instance, the Berry curvature dipole (BCD)~\cite{Sodemann2015} in noncentrosymmetric metals can mediate a linear EO Pockels effect~\cite{Konig2019,Cheng2019,Li2022,Shi2023,Rappoport2023}. Activated when rotational symmetry in 2D noncentrosymmetric metals is broken, this effect, however, is weak. It often requires highly engineered platforms such as strained flat band materials e.g., strained twisted bilayer graphene~\cite{Rappoport2023,Pantaleon2021,He2021}. Similarly, while the non Drude-like dynamics of carriers (e.g., arising from third-order derivative of the velocity) can mediate an EO Kerr effect~\cite{Almasov1971}, such Kerr effects require highly dispersive bands -- vanishing for linear, parabolic, and cubic electronic dispersion relations. As a result, pronounced metallic EO effects can be challenging to realize and control.

\begin{table*}
\begin{center}
\begin{tabular}{l c c c c}
\hline\hline
 &                  & dependence on $\tau$  & $\mathrm{Re}\delta\sigma$ dependence on $\omega$ & $\mathrm{Im}\delta\sigma$ dependence on $\omega$  \\
 & $C_{3v}$ & ($\omega \to 0$) & ($\omega\tau > 1$) & ($\omega\tau > 1$) \\ 
\hline 
$\sigma_{0}$ & $+$ & $\tau$ & $\frac{1}{\omega^{2}}$ & $\frac{1}{\omega}$ \\
\hline 
Pockels Effect \\
\hline 
Berry curvature dipole (BCD) (antisymmetric) & $-$ & $\tau$ & 1 & 1 \\
Berry curvature dipole (BCD) (non-Hermitian) & $-$ & $\tau$ & $\frac{1}{\omega^{2}}$ & $\frac{1}{\omega}$\\
Skew scattering & $+$ & $\tau^{3}$ & $\frac{1}{\omega^{2}}$ & $\frac{1}{\omega}$ \\
\hline 
Nonlinear Kerr Effect \\
\hline 
``Snap'' Kerr effect & $+$ & $\tau^{3}$  & $\frac{1}{\omega^{4}}$ & $\frac{1}{\omega}$ \\ 
\hline \hline
\end{tabular}
\end{center}
\caption{Symmetry, $\tau$ dependence and $\omega$ dependence of various sources of the Pockels and the Kerr effects. $+$ means that the relevant mechanisms are allowed by $C_{3v}$ point group symmetry in a 2D system, while $-$ means they are forbidden. The antisymmetric part of the BCD Pockels does not depend on $\omega$, shown as $1$ in the table.}
\label{Table:Mechanisms}
\end{table*}

Here we find that large metallic EO can be naturally realized in gapped 2D materials without strain or flat electronic bands. In particular, we argue that skew-scattering of carriers in a metal yields a Pockels EO effect at terahertz frequencies. Different from the BCD-induced EO effect~\cite{Konig2019,Cheng2019,Li2022,Shi2023,Rappoport2023}, the skew scattering Pockels effect survives rotational symmetry, yielding large pronounced values that dominate the metallic EO in clean noncentrosymmertic metals. As we will see below, the high mobility readily achieved for 2D materials such as gapped bilayer graphene lead to pronounced EO Pockels effect at relatively small applied DC electric fields.

Strikingly, beyond the skew-scattering Pockels EO effect, we find that the high-order dispersions induced in gapped bilayer graphene also provide a natural venue for a ``Snap'' Kerr effect. As we explain, both these skew-scattering Pockels and ``Snap'' Kerr EO effects compete in gapped bilayer graphene to produce a rich range of metallic EO-induced phenomena such as a field-induced birefringence that is non-reciprocal in applied DC field as well as field-modulated transmission and reflectance. Such 2D metallic EO effects provide a natural and readily accessible platform to control a full range of electromagnetic properties at terahertz frequencies.

\begin{figure}
\centering
\includegraphics[width=\linewidth]{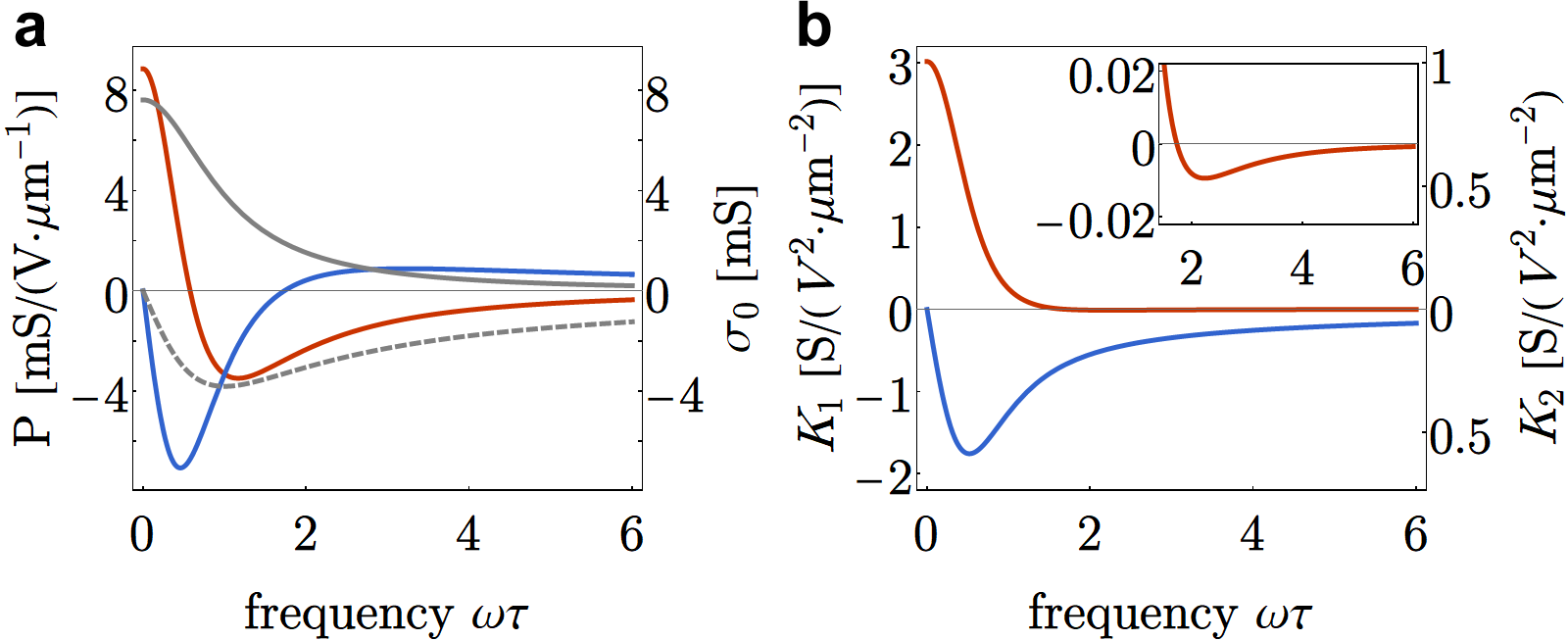}
\caption{Metallic Pockels and Kerr coefficients in gapped bilayer graphene. (a) Real (red) and imaginary (blue) parts of the skew-scattering Pockels coefficient $P$, and real (soild gray) and imaginary (dashed gray) part of the field-free optical conductivity $\sigma_{0}$. (b) Real (red) and imaginary (blue) part of the ``Snap'' Kerr coefficients $K_{1}$ (left axis) and $K_{2}$ (right axis). Note that at about $\omega\tau > 1$ the real and the imaginary parts of $\sigma_{0}$, $P$, $K_{1}$ and $K_{2}$ all decay with increasing frequency, but the rate of decay is not the same, as shown in Table.~\ref{Table:Mechanisms}. Here we use $\lambda=1.754 \times 10^{-1} \ \mathrm{m/s}$, $v_{t} = 1 \times 10^{5} \ \mathrm{m/s}$, $\Delta= 50\ \mathrm{meV}$, the impurity density $n_{i}= 1 \times 10^{9} \ \mathrm{cm}^{-2}$, the impurity potential $V_{0}= 1 \times 10^{-13}\ \mathrm{eV}\cdot \mathrm{cm}^{2}$~\cite{Liang2020} and chemical potential $\mu=65\ \mathrm{meV}$.}
\label{fig:P-and-K-coefficients}
\end{figure}

{\it Mechanisms of the electro-optic effect in metals:} We begin by systematically characterizing the mechanisms for metallic EO. The dynamical conductivity, $\sigma_{\alpha\beta}  \left( \omega \right)$, can be modified by an applied DC electric field ${\bf E}$, $\sigma_{\alpha\beta}  \left( \omega \right) = \sigma^{(0)}_{\alpha\beta}  \left( \omega \right) + \delta\sigma_{\alpha\beta}  \left( \omega \right)$, where $\sigma^{(0)} \left( \omega \right)$ is the bare (field-free) optical conductivity and $\delta \sigma_{\alpha \beta}$ describes the electro-optic contribution
\begin{equation}
\delta\sigma_{\alpha\beta}  \left( \omega \right) = P_{\alpha\beta\gamma} E_{\gamma}+ K_{\alpha\beta\gamma\delta} E_{\gamma} E_{\delta} + \mathcal{O}(E^3) 
\label{eq:EO}
\end{equation}
where $\{\alpha, \beta, \gamma\} = \{x,y,z\}$ indices are Cartesian coordinates. The linear in $E$ part is termed the Pockels effect ($P$), and the quadratic in $E$ part is referred to as the nonlinear Kerr effect ($K$). The Pockels effect requires broken centrosymmetry while the nonlinear Kerr effect does not.

To analyze various contribution to the electro-optic effect in metals, we examine the charge current dynamics in the electronic band of a metal. These can be described semiclassically as 
\begin{equation}
\vec j (t) =  - e \sum_{\vec k} \big[\vec{v} (\vec k) + e \boldsymbol{\mathcal{E}} (t)/\hbar \times  \boldsymbol{\Omega} (\vec k) \big] f (\vec k,t),
\label{eq:current} 
\end{equation}
where the electron charge is $-e<0$, $\vec k$ is the wavevector of a Bloch electron, and $f (\vec k,t)$ is the electronic distribution function. Here $\vec{v} (\vec k) = \partial \epsilon (\vec k)/\hbar \partial \vec k $ is the electronic group velocity [with electronic dispersion $\epsilon (\vec k)$], $\boldsymbol{\Omega} (\vec k)$ is the Berry curvature, and $\boldsymbol{\mathcal{E}} (t) = \boldsymbol{\mathcal{E}}_{\rm AC} (t) + \vec E$ is the total electric field that includes both an oscillating part $\boldsymbol{\mathcal{E}}_{AC}(t) = \boldsymbol{\mathcal{E}}^{(0)} e^{i\omega t} + c.c.$ as well as an applied DC field, $\vec E$, that produces the electro-optic effects we unveil. 

The distribution function in Eq.~(\ref{eq:current}) can be obtained by analyzing the spatially uniform Boltzmann equation 
\begin{equation}
\partial_t f (\vec k, t) - e \boldsymbol{\mathcal{E}} (t) \cdot \partial_{\vec k} f (\vec k, t) / \hbar = I \{ f(\vec k, t) \}, 
\end{equation} 
where $I \{ f(\vec k, t) \}$ is the collision integral that can include both symmetric (with rate $w^{\mathrm{S}}_{{\bf k}{\bf k}^{\prime}}$) as well as anti-symmetric (skew) scattering (with rate $w^{\mathrm{A}}_{{\bf k}{\bf k}^{\prime}}$). 

We solve the distribution function iteratively in the standard perturbative fashion [in powers of $\mathcal{E}(t)$] with the relaxation time approximation, see {\bf SI}~\cite{SeeSI} for a detailed discussion. For example, in the absence of applied DC electric fields, $\vec E=0$, the bare optical conductivity reads 
\be
\sigma^{(0)}_{\alpha\beta} (\omega) = - e^{2} \tau_\omega \sum_{\vec k} v_\alpha v_\beta \frac{ \partial f_0 (\vec k)}{\partial \epsilon} , \quad \tau_\omega = \frac{\tau}{1+ i \omega \tau}
\label{eq:bare}
\ee 
where $1/\tau= \sum_{\vec k} w^{\mathrm{S}}_{{\bf k}{\bf k}^{\prime}} (1-{\rm cos} \theta_{\vec k, \vec k'})$, $f_0 (\vec k)$ is the equilibrium Fermi distribution, and $\theta_{\vec k, \vec k'}$ is the angle between $\vec{v} ({\bf k}^{\prime})$ and $\vec{v} (\vec k)$. Eq.~(\ref{eq:bare}) describes the conventional Drude dynamics characterized by a frequency dependence that goes as $\tau_\omega$. 

When a DC electric field $\vec E \neq 0$ is applied, we find the current dynamics in Eq.~(\ref{eq:current}) can change qualitatively yielding (non-Drude) frequency dependence that departs from $\tau_\omega$. As we will see, these produce EO effects in Eq.~(\ref{eq:EO}). Extracting the optical conductivity from Eq.~(\ref{eq:current}) and keeping terms up to the quadratic order of the applied DC electric field, we identify three mechanisms of EO of non-magnetic metals summarized in Table~\ref{Table:Mechanisms} (full details of derivation can be found in the {\bf SI}~\cite{SeeSI}). 

The Pockels effect has two major origins, $P_{\alpha\beta\gamma} = P_{\alpha\beta\gamma}^{(\mathrm{BCD})} + P_{\alpha\beta\gamma}^{(\mathrm{skew})}.$ The first is the BCD Pockels effect which arises from combining the anomalous velocity induced by Berry curvature and distribution function with symmetric scattering~\cite{Konig2019,Cheng2019,Li2022,Shi2023,Rappoport2023} 
\begin{equation}
P_{\alpha\beta\gamma}^{(\mathrm{BCD})} = \frac{e^{3}}{\hbar^{2}} \left( \tau_{\omega} \varepsilon_{\alpha \gamma \delta} D_{\beta \delta} + \tau \varepsilon_{\alpha \beta \delta}D_{\gamma \delta}  \right), 
\label{eq:BCD}
\end{equation}
where $D_{\alpha \beta} = \sum_{{\bf k}} f_{0} \left({\bf k} \right) [\partial \Omega_{\beta} /\partial k_{\alpha}]$ is the Berry curvature dipole (BCD)~\cite{Sodemann2015,QiongMa2018,MakKinFai2019}, a quantum geometric quantity. $P_{\alpha\beta\gamma}^{(\mathrm{BCD})}$ consists of two parts. The first is non-Hermitian: it alters energy absorption of the metal and may even contribute to amplification~\cite{Rappoport2023,Shi2023}. In contrast, the other term in Eq.~(\ref{eq:BCD}) is purely antisymmetric, i.e., Hall in nature. It does not do any work, and it does not affect energy absorption. It is, however, insensitive to frequency and dominates the high frequency response of $P_{\alpha\beta\gamma}^{(\mathrm{BCD})}$ producing a gyrotropic Hall effect that rotates the polarization of incident linearly polarized light~\cite{Konig2019}. 

The other Pockels effect, which is one of the main results of our work, arises from dynamics of the distribution function induced by skew-scattering: 
\begin{align}
P_{\alpha\beta\gamma}^{(\mathrm{skew})} &= \frac{e^{3}}{\hbar^{2}} \left\{ \tau_{\omega}^{3} S^{(1)}_{\alpha \gamma \beta } + \tau_{\omega} \tau^{2} S^{(1)}_{\alpha \beta \gamma } + \tau_{\omega}^{2} \left( \tau_{\omega} + \tau \right) S^{(2)}_{\alpha \beta \gamma } \right\} , \nonumber \\
S^{(1)}_{\alpha \beta \gamma } &= - \frac{1}{\hbar} \sum_{{\bf k}} \frac{\partial \epsilon }{\partial k_{\alpha}}  \frac{\partial  }{\partial k_{\beta}} \sum_{{\bf k}^{\prime}} w^{\mathrm{A}}_{{\bf k}{\bf k}^{\prime}} \frac{\partial  }{\partial k^{\prime}_{\gamma}} f_{0} \left({\bf k}^{\prime} \right), \nonumber \\
S^{(2)}_{\alpha \beta \gamma } &= - \frac{1}{\hbar} \sum_{{\bf k}} \frac{\partial \epsilon }{\partial k_{\alpha}} \sum_{{\bf k}^{\prime}} w^{\mathrm{A}}_{{\bf k}{\bf k}^{\prime}}  \frac{\partial^{2}  }{\partial k^{\prime}_{\beta} \partial k^{\prime}_{\gamma}} f_{0} \left({\bf k}^{\prime} \right).
\label{eq:Pockels-Skew}
\end{align} 
The skew-scattering Pockels effect is a quatum geometric effect, too, as the skew scattering rate $w^{\mathrm{A}}_{{\bf k}{\bf k}^{\prime}} $ is related to the Pancharatnam phase~\cite{Sinitsyn2006}. We note that the antisymmetric part of $P_{\alpha\beta\gamma}^{(\mathrm{skew})}$ also contributes to the gyrotropic Hall effect~\cite{Konig2019} but vanishes in rotationally symmetric systems. As we will see below, $P_{\alpha\beta\gamma}^{(\mathrm{skew})}$ can also be symmetric, persisting in rotationally symmetric system, and provides the dominant Pockels effect in clean metallic systems.

\begin{figure}
\centering
\includegraphics[width=\linewidth]{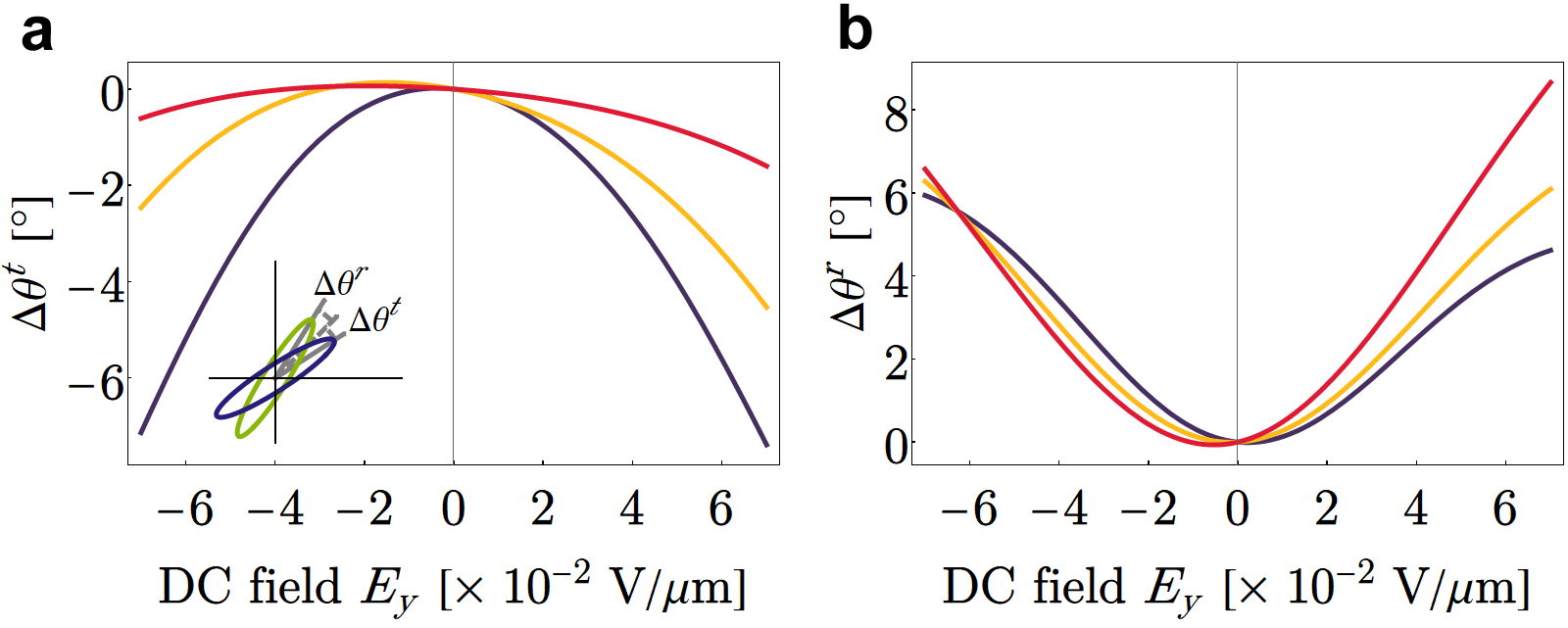}
\caption{Field induced birefringence and polarization rotation induced by metallic EO in gapped bilayer graphene. (a) Polarization rotation for transmitted light controlled by in-plane DC electric field. (inset) The incident light is polarized at an angle of $\pi/4$ to the $x$ axis, and the reflected (green) and the transmitted (blue) light are in general elliptically-polarized with major axes rotated by (a) $\Delta \theta^{t}$ and (b) $\Delta \theta^{r}$, respectively. (b) Polarization rotation for reflected light controlled by in-plane DC electric field. Both skew-scattering Pockels EO and ``Snap'' Kerr EO contribute to the polarization rotation. Their competition leads to polarization rotation that is non-reciprocal in DC electric field. Note that the linear region (i.e. dominated by Pockels EO) grows as $\omega\tau$ increases. Purple, orange and red curves correspond to $\omega\tau = 1, 2, 5$.  Bilayer graphene parameters same as Fig.~\ref{fig:P-and-K-coefficients}. }
\label{fig:Polarization-Rotation}
\end{figure}

Lastly, we find the main contribution to the Kerr effect in clean metals is a ``Snap'' term that arises from the non-Drude dynamics of carriers. This arises from the conventional velocity and high-order electric field induced changes to the distribution function from symmetric scattering,
\begin{align}
\left[ K_{\alpha\beta}^{\gamma\delta} \right]_{\mathrm{Snap}} &= \left(  \tau_{\omega}^{3} +  \tau_{\omega}^{2} \tau +  \tau_{\omega} \tau^{2} \right) J_{\alpha\beta \gamma \delta}, \nonumber \\
J_{\alpha\beta \gamma \delta} &= - \frac{e^{4}}{\hbar^{4}} \sum_{{\bf k}}  \frac{\partial^{3} \epsilon }{\partial k_{\alpha} \partial k_{\beta} \partial k_{\gamma} } \frac{\partial }{ \partial k_{\delta}} f_{0} \left({\bf k} \right) .
\label{eq:Kerr-Drude}
\end{align} 
It is not geometric, as it is determined by band dispersion only. $\left[ K_{\alpha\beta}^{\gamma\delta} \right]_{\mathrm{Snap}}$ gives a symmetric contribution to the optical conductivity, since $J_{\alpha\beta \gamma \delta}$ is invariant under permutation of indices. Importantly, $J_{\alpha\beta \gamma \delta}$ effectively depends on a ``snap'' (also called Jounce)~\cite{Sprott1997}, a third-order derivative of the velocity. As a result, it requires bands to possess high-order dispersions~\cite{Almasov1971}: $J_{\alpha\beta \gamma \delta}$ vanishes for linear, parabolic and cubic dispersion relations, rendering its effects challenging to realize. As we will see below, the non-parabolic bands of gapped bilayer graphene provides an ideal venue to activate $J_{\alpha\beta \gamma \delta}$. 

We note, parenthetically, that apart from the ``Snap'', another mechanism could also contribute to the Kerr effect: the Berry connection polarizability (BCP)~\cite{Gao2014}. This contribution can be obtained from Eq.~(\ref{eq:current}) by taking into account the BCP correction to the velocity and Berry curvature. However, in clean systems with long relaxation time $\tau$, the Snap Kerr effect readily dominates over it. As a result, here we will focus on Snap contributions. 

These EO mechanisms have different dependences on relaxation time and frequency, summarized in Table~\ref{Table:Mechanisms}. For instance, at low frequencies, the BCD Pockels contribution is proportional to $\tau$, while the skew-scattering Pockels and the ``Snap'' Kerr are proportional to $\tau^{3}$. This renders skew-scattering Pockels and ``Snap'' Kerr effects the dominant EO effects in clean metals. Interestingly, as $\omega$ increases, the real part of the skew-scattering Pockels and the ``Snap'' Kerr effects scale as $\frac{1}{\omega^{2}}$ and $\frac{1}{\omega^{4}}$ at $\omega\tau > 1$ respectively; their imaginary parts scale in the same way. As a result, as we will see below, skew-scattering Pockels effects play a dominant role at high frequencies.  

{\it The electro-optic effect in gapped bilayer graphene:}\ To illustrate the intraband metallic EO, we examine gapped bilayer graphene that can be described via the effective Hamiltonian~\cite{Liang2020}:
\begin{align}
\label{eq:BLG}
H_{s} \left( {\bf k} \right) = & \left\{ s \hbar v_{t} k_{x} - \hbar \lambda \left( k_{x}^{2} - k_{y}^{2} \right) \right\} \sigma_{x} \nonumber \\
& + \left( \hbar v_{t} k_{y} + 2 s \hbar \lambda k_{x} k_{y} \right) \sigma_{y} + \Delta \sigma_{z},
\end{align}
where $s=\pm 1$ represents valley degree of freedom, $\lambda$ characterizes the quadratic band dispersion, $v_{t}$ describes trigonal warping, $\Delta$ is the half gap, and $\sigma$ matrices correspond to the sublattice degree of freedom. In what follows, we will focus on the clean limit readily realized in gapped bilayer graphene with relaxation time $\tau \approx 1 \ \mathrm{ps}$. 

Gapped bilayer graphene possesses a $C_{3v}$ point group symmetry with one mirror plane along the $y$-axis so that $H_{s} \left( {\bf k} \right)$ is invariant under $k_{x} \leftrightarrow -k_{x}$ and $s \leftrightarrow -s$. $C_{3v}$ zeros the BCD~\cite{Sodemann2015} while allowing symmetric skew-scattering Pockels and ``Snap'' Kerr effects to survive. $C_{3v}$ also ensures that the field-free optical conductivity $\sigma^{(0)} \left( \omega \right)$ is isotropic. We find the EO contribution to the optical conductvitiy tensor, $\delta\sigma(\omega)$, is symmetric and is described by 
\begin{equation}
\label{eq:delta-sigma}
\delta\sigma = \left( 
\begin{array}{cc}
P E_{y} + K_{1} E_{x}^{2} + K_{2} E_{y}^{2} & P E_{x} + \Delta K E_{x} E_{y}\\
P E_{x} +  \Delta K E_{x} E_{y} & -P E_{y} + K_{1} E_{y}^{2} + K_{2} E_{x}^{2}\\
\end{array}
\right),
\end{equation}
where the Pockels and the Kerr tensors in Eq.~(\ref{eq:EO}), $P_{\alpha\beta\gamma}$ and $K_{\alpha\beta\gamma\delta}$, are reduced to three numbers: $P$ for Pockels, and $K_{1}$ and $K_{2}$ for Kerr, with $\Delta K = \left( K_{1} - K_{2} \right)$. Interestingly, we note that $C_{3v}$ symmetry dictates that the ``Snap'' Kerr contribution $K_{1} = 3K_{2}$. 

Using Eqs.~(\ref{eq:Pockels-Skew}) and (\ref{eq:Kerr-Drude}) and the Hamiltonian in Eq.~(\ref{eq:BLG}), we find pronounced intraband EO coefficients characterized by P and K coefficients in Fig.~\ref{fig:P-and-K-coefficients}. Indeed, small applied DC $\vec E \sim 0.02 {\rm V}/ \mu {\rm m}$ readily produce EO $\delta\sigma$ of order 10 $\%$ of the field-free bare conductivity $\sigma_0(\omega)$ [gray curves]. As expected, ${\rm Re}K$ rapidly decreases for large frequencies allowing ${\rm Re} P$ to dominate the real part (responsible for absorptive effects) of EO for $\omega\tau > 1$. In contrast, both ${\rm Im}K$ and ${\rm Im} P$ decay as $1/\omega$ at large frequencies; given the large intrinsic values of ${\rm Im} K$, the Kerr effect readily dominates the imaginary part of EO for modest values of applied $\vec E$. As a result, as we will see below, the interplay of both Pockels and Kerr effects are critical in describing EO related phenomena such as field-induced birefringence and field-tunable reflectance and transmission.

The pronounced intraband EO coefficients enable to modulate the optical properties of the Fermi surface. To see this, we first examine the reflectance and transmission tensors $\bar{r}$ and $\bar{t}$ linking 
the incident ($\boldsymbol{\mathcal{E}}^{i}$), the reflected ($\boldsymbol{\mathcal{E}}^{\mathrm{r}}$) and the transmitted ($\boldsymbol{\mathcal{E}}^{\mathrm{t}}$) electric fields as $\left( {\mathcal{E}}^{\mathrm{r}}_{x}, {\mathcal{E}}^{\mathrm{r}}_{y} \right)^{\mathrm{T}} = \bar{r} \left( {\mathcal{E}}^{i}_{x}, {\mathcal{E}}^{i}_{y} \right)^{\mathrm{T}}$ and $\left( {\mathcal{E}}^{\mathrm{t}}_{x}, {\mathcal{E}}^{\mathrm{t}}_{y} \right)^{\mathrm{T}} = \bar{t} \left( {\mathcal{E}}^{i}_{x}, {\mathcal{E}}^{i}_{y} \right)^{\mathrm{T}}$. We solve for $\bar{r}$ and $\bar{t}$ across a thin film via the standard electromagnetic boundary conditions, see {\bf SI}~\cite{SeeSI}. 

\begin{figure}
\centering
\includegraphics[width=\linewidth]{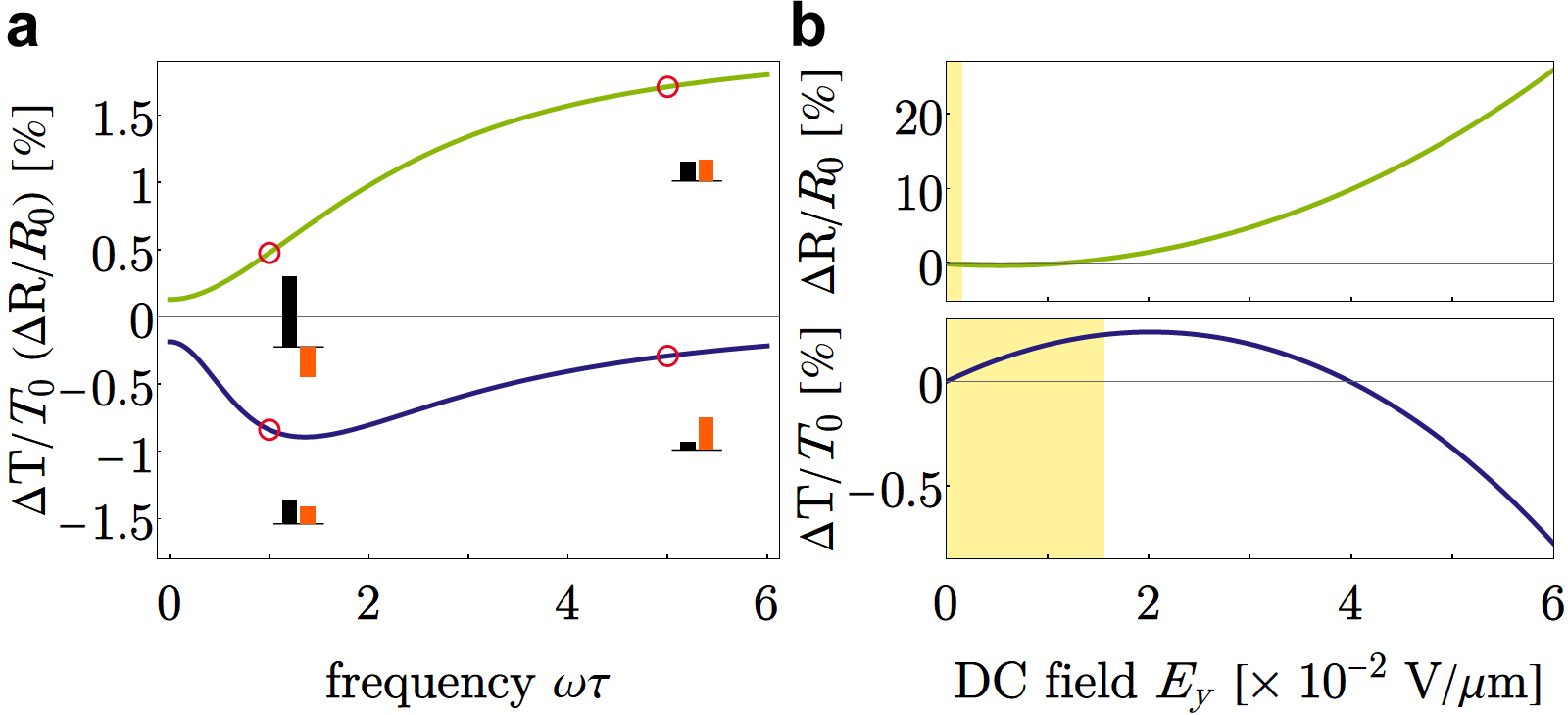}
\caption{Metallic EO effect modulated transmission and reflectance in gapped bilayer graphene. (a) EO modulation of the transmittance and reflectance at $E_{y} = -0.01 \ \mathrm{V/\mu m}$ as a percentage of the field-free transmittance (blue, $\Delta T/T_{0}$) and reflectance (green, $\Delta R/R_{0}$). The bars show the relative contributions of $\Delta T/T_{0}$ or $\Delta R/R_{0}$ contributed by skew-scattering Pockels (orange) and ``Snap'' Kerr (black) at frequencies highlighted by circles, $\omega\tau =1$ and $5$. (b) $\Delta T/T_{0}$ (blue, bottom panel) and $\Delta R/R_{0}$ (green, top panel) as a function of applied DC electric field $E_{y}$ at $\omega\tau = 5$. Both curves have a linear behavior at small $E_{y}$ indicating Pockels dominance (areas in orange) and deviating from linear trend at higher $E_{y}$ as the Kerr effect kicks in. The other parameters are the same as in Fig.~\ref{fig:P-and-K-coefficients}. }
\label{fig:reflectance-and-transmittance}
\end{figure}

We first note that the bare field-free conductivity is isotropic. However, an applied $\vec E$ breaks this isotropy manifesting a field-induced birefringence yielding a polarization rotation for an initially linearly polarized beam. As an illustration, we consider applied $\vec E$ along the $y$-axis and incident light with a polarization aligned $\pi/4$ away from the $y$-axis. We find the intraband EO effects in Eq.~(\ref{eq:delta-sigma}) produce a polarization rotation for both transmitted and reflected beams, see Fig.~\ref{fig:Polarization-Rotation}. Interestingly, even as the incident light is linearly polarized, both transmitted and reflected beams are elliptically polarized. This is because $P$, $K_{1}$ and $K_{2}$ in general are not in-phase. As a result, for incident beams not along principal axes, $\delta\sigma$ cannot be diagonalized through unitary transformations.

The field-induced polarization rotation shown in Fig.~\ref{fig:Polarization-Rotation} is non-monotonic as a function of $\vec E$ and is non-reciprocal: i.e., polarization rotation is distinct for $+\vec E$ vs $-\vec E$. This behavior arises from a competition of both Kerr and Pockels effects (even vs odd in $\vec E$) yielding a displaced dome/valley-like behavior for the polarization rotation that shifts with frequency. The region of applied $E_y$ where Pockels EO dominates also increases in agreement with the $\omega$ dependence of $P$ and $K$ discussed above.

We can also track the intraband EO effect by examining how the intensity of transmitted and reflected light can be modulated by the applied $\vec E$ through an EO induced differential transmittance and reflectance, plotted in Fig.~\ref{fig:reflectance-and-transmittance}. Here $\Delta T(\Delta R) =  T_{\rm total}(R_{\rm total}) - T_0(R_0)$ where the subscript $0$ denotes the bare transmitted/reflected intensity. In the same fashion as in the polarization rotation, both Pockels and Kerr effects contribute to the EO induced differential transmittance and reflectance (see bar insets for comparison of relative magnitude). However, their contributions are differentiated for reflectance and transmittance. At high frequencies, both Pockels and Kerr effects play comparable roles in the reflectance; in contrast, the Pockels effect dominates the transmittance (Fig.~\ref{fig:reflectance-and-transmittance}a). Indeed, the range of $\vec E$ values where Pockels effects dominates are distinct for transmittance and reflectance (Fig.~\ref{fig:reflectance-and-transmittance}b); Pockels dominates for a larger range of $\vec E$ values for transmittance, while reflectance very quickly enters into a Kerr dominated regime. Strikingly, we find EO induced changes can be sizeable reaching ratios of order 10$\%$ even for modest fields of $\vec E = 0.05 \, {\rm V}/\mu {\rm m}$ (Fig.~\ref{fig:reflectance-and-transmittance}b).

The scattering processes in a metal can lead to sizeable and a varied set of intraband EO effects, including the skew-scattering Pockels effect and ``Snap'' Kerr effect that dominate the clean limit. Different from EO effects in insulators, intraband EO effects operate at low frequencies, with a characteristic scale $\omega \sim 1/\tau$; for clean materials such as gapped bilayer graphene, $\tau \sim 1\, {\rm ps}$ yielding characteristic operating frequencies in the terahertz. While we have focussed on gapped bilayer graphene (with sizeable EO effects), we anticipate the intraband EO effects that are enhanced at long scattering times discussed here readily apply to a wide range of materials. These mechanisms may add to and augment a growing class of opto-electronic devices that rely on EO effects of the Fermi surface~\cite{Konig2019,Cheng2019,Li2022,Shi2023,Rappoport2023}.

{\it Acknowledgement:}  This work was supported by Singapore Ministry of Education (MOE) AcRF Tier 2 grant MOE-T2EP50222-0011 and MOE Tier 3 grant MOE 2018-T3-1-002.

\bibliographystyle{apsrev4-1}
\bibliography{Bib}

\begin{thebibliography}{31}%
\makeatletter
\providecommand \@ifxundefined [1]{%
 \@ifx{#1\undefined}
}%
\providecommand \@ifnum [1]{%
 \ifnum #1\expandafter \@firstoftwo
 \else \expandafter \@secondoftwo
 \fi
}%
\providecommand \@ifx [1]{%
 \ifx #1\expandafter \@firstoftwo
 \else \expandafter \@secondoftwo
 \fi
}%
\providecommand \natexlab [1]{#1}%
\providecommand \enquote  [1]{``#1''}%
\providecommand \bibnamefont  [1]{#1}%
\providecommand \bibfnamefont [1]{#1}%
\providecommand \citenamefont [1]{#1}%
\providecommand \href@noop [0]{\@secondoftwo}%
\providecommand \href [0]{\begingroup \@sanitize@url \@href}%
\providecommand \@href[1]{\@@startlink{#1}\@@href}%
\providecommand \@@href[1]{\endgroup#1\@@endlink}%
\providecommand \@sanitize@url [0]{\catcode `\\12\catcode `\$12\catcode
  `\&12\catcode `\#12\catcode `\^12\catcode `\_12\catcode `\%12\relax}%
\providecommand \@@startlink[1]{}%
\providecommand \@@endlink[0]{}%
\providecommand \url  [0]{\begingroup\@sanitize@url \@url }%
\providecommand \@url [1]{\endgroup\@href {#1}{\urlprefix }}%
\providecommand \urlprefix  [0]{URL }%
\providecommand \Eprint [0]{\href }%
\providecommand \doibase [0]{http://dx.doi.org/}%
\providecommand \selectlanguage [0]{\@gobble}%
\providecommand \bibinfo  [0]{\@secondoftwo}%
\providecommand \bibfield  [0]{\@secondoftwo}%
\providecommand \translation [1]{[#1]}%
\providecommand \BibitemOpen [0]{}%
\providecommand \bibitemStop [0]{}%
\providecommand \bibitemNoStop [0]{.\EOS\space}%
\providecommand \EOS [0]{\spacefactor3000\relax}%
\providecommand \BibitemShut  [1]{\csname bibitem#1\endcsname}%
\let\auto@bib@innerbib\@empty
\bibitem [{\citenamefont {Shen}(1984)}]{Shen}%
  \BibitemOpen
  \bibfield  {author} {\bibinfo {author} {\bibfnamefont {Y.~R.}\ \bibnamefont
  {Shen}},\ }\href@noop {} {\emph {\bibinfo {title} {The Principles of
  Nonlinear Optics}}}\ (\bibinfo  {publisher} {Wiley Interscience},\ \bibinfo
  {address} {New York},\ \bibinfo {year} {1984})\BibitemShut {NoStop}%
\bibitem [{\citenamefont {Jenkins}\ and\ \citenamefont
  {White}(2001)}]{Jenkins}%
  \BibitemOpen
  \bibfield  {author} {\bibinfo {author} {\bibfnamefont {F.~A.}\ \bibnamefont
  {Jenkins}}\ and\ \bibinfo {author} {\bibfnamefont {H.~E.}\ \bibnamefont
  {White}},\ }\href@noop {} {\emph {\bibinfo {title} {Fundamentals of
  Optics}}},\ \bibinfo {edition} {4th}\ ed.\ (\bibinfo  {publisher}
  {McGraw-Hill},\ \bibinfo {address} {New York},\ \bibinfo {year}
  {2001})\BibitemShut {NoStop}%
\bibitem [{\citenamefont {New}(2011)}]{New}%
  \BibitemOpen
  \bibfield  {author} {\bibinfo {author} {\bibfnamefont {G.}~\bibnamefont
  {New}},\ }\href {\doibase 10.1017/CBO9780511975851} {\emph {\bibinfo {title}
  {Introduction to nonlinear optics}}}\ (\bibinfo  {publisher} {Cambridge
  University Press},\ \bibinfo {address} {Cambridge},\ \bibinfo {year}
  {2011})\BibitemShut {NoStop}%
\bibitem [{\citenamefont {Vorob'ev}\ \emph {et~al.}(1979)\citenamefont
  {Vorob'ev}, \citenamefont {Ivchenko}, \citenamefont {Pikus}, \citenamefont
  {Farbshtein}, \citenamefont {Shalygin},\ and\ \citenamefont
  {Shturbin}}]{Vorobev1979}%
  \BibitemOpen
  \bibfield  {author} {\bibinfo {author} {\bibfnamefont {L.~E.}\ \bibnamefont
  {Vorob'ev}}, \bibinfo {author} {\bibfnamefont {E.~L.}\ \bibnamefont
  {Ivchenko}}, \bibinfo {author} {\bibfnamefont {G.~E.}\ \bibnamefont {Pikus}},
  \bibinfo {author} {\bibfnamefont {I.~I.}\ \bibnamefont {Farbshtein}},
  \bibinfo {author} {\bibfnamefont {V.~A.}\ \bibnamefont {Shalygin}}, \ and\
  \bibinfo {author} {\bibfnamefont {A.~V.}\ \bibnamefont {Shturbin}},\ }\href
  {http://jetpletters.ru/ps/0/article_22128.shtml} {\bibfield  {journal}
  {\bibinfo  {journal} {JETP Lett.}\ }\textbf {\bibinfo {volume} {29}},\
  \bibinfo {pages} {441} (\bibinfo {year} {1979})}\BibitemShut {NoStop}%
\bibitem [{\citenamefont {Shalygin}(2022)}]{Shalygin2022}%
  \BibitemOpen
  \bibfield  {author} {\bibinfo {author} {\bibfnamefont {V.~A.}\ \bibnamefont
  {Shalygin}},\ }in\ \href {\doibase 10.1007/978-3-031-11287-4_1} {\emph
  {\bibinfo {booktitle} {Optics and Its Applications}}},\ \bibinfo {editor}
  {edited by\ \bibinfo {editor} {\bibfnamefont {D.}~\bibnamefont {Blaschke}},
  \bibinfo {editor} {\bibfnamefont {D.}~\bibnamefont {Firsov}}, \bibinfo
  {editor} {\bibfnamefont {A.}~\bibnamefont {Papoyan}}, \ and\ \bibinfo
  {editor} {\bibfnamefont {H.~A.}\ \bibnamefont {Sarkisyan}}}\ (\bibinfo
  {publisher} {Springer International Publishing},\ \bibinfo {address} {Cham},\
  \bibinfo {year} {2022})\ pp.\ \bibinfo {pages} {1--19}\BibitemShut {NoStop}%
\bibitem [{\citenamefont {Shi}\ \emph {et~al.}(2023)\citenamefont {Shi},
  \citenamefont {Matsyshyn}, \citenamefont {Song},\ and\ \citenamefont
  {Villadiego}}]{Shi2023}%
  \BibitemOpen
  \bibfield  {author} {\bibinfo {author} {\bibfnamefont {L.-k.}\ \bibnamefont
  {Shi}}, \bibinfo {author} {\bibfnamefont {O.}~\bibnamefont {Matsyshyn}},
  \bibinfo {author} {\bibfnamefont {J.~C.~W.}\ \bibnamefont {Song}}, \ and\
  \bibinfo {author} {\bibfnamefont {I.~S.}\ \bibnamefont {Villadiego}},\ }\href
  {\doibase 10.1103/PhysRevB.107.125151} {\bibfield  {journal} {\bibinfo
  {journal} {Phys. Rev. B}\ }\textbf {\bibinfo {volume} {107}},\ \bibinfo
  {pages} {125151} (\bibinfo {year} {2023})}\BibitemShut {NoStop}%
\bibitem [{\citenamefont {Rappoport}\ \emph {et~al.}(2023)\citenamefont
  {Rappoport}, \citenamefont {Morgado}, \citenamefont {Lanneb\`ere},\ and\
  \citenamefont {Silveirinha}}]{Rappoport2023}%
  \BibitemOpen
  \bibfield  {author} {\bibinfo {author} {\bibfnamefont {T.~G.}\ \bibnamefont
  {Rappoport}}, \bibinfo {author} {\bibfnamefont {T.~A.}\ \bibnamefont
  {Morgado}}, \bibinfo {author} {\bibfnamefont {S.}~\bibnamefont
  {Lanneb\`ere}}, \ and\ \bibinfo {author} {\bibfnamefont {M.~G.}\ \bibnamefont
  {Silveirinha}},\ }\href {\doibase 10.1103/PhysRevLett.130.076901} {\bibfield
  {journal} {\bibinfo  {journal} {Phys. Rev. Lett.}\ }\textbf {\bibinfo
  {volume} {130}},\ \bibinfo {pages} {076901} (\bibinfo {year}
  {2023})}\BibitemShut {NoStop}%
\bibitem [{\citenamefont {Hakimi}\ \emph {et~al.}()\citenamefont {Hakimi},
  \citenamefont {Rouhi}, \citenamefont {Rappoport}, \citenamefont
  {Silveirinha},\ and\ \citenamefont {Capolino}}]{Hakimi2023}%
  \BibitemOpen
  \bibfield  {author} {\bibinfo {author} {\bibfnamefont {A.}~\bibnamefont
  {Hakimi}}, \bibinfo {author} {\bibfnamefont {K.}~\bibnamefont {Rouhi}},
  \bibinfo {author} {\bibfnamefont {T.~G.}\ \bibnamefont {Rappoport}}, \bibinfo
  {author} {\bibfnamefont {M.~G.}\ \bibnamefont {Silveirinha}}, \ and\ \bibinfo
  {author} {\bibfnamefont {F.}~\bibnamefont {Capolino}},\ }\href@noop {}
  {}\bibinfo {note} {Preprint at \url{http://arXiv.org/abs/2312.15142}
  (2023)}\BibitemShut {NoStop}%
\bibitem [{\citenamefont {Morgado}\ \emph {et~al.}()\citenamefont {Morgado},
  \citenamefont {Rappoport}, \citenamefont {Tsirkin}, \citenamefont
  {Lannebère}, \citenamefont {Souza},\ and\ \citenamefont
  {Silveirinha}}]{Morgado2024}%
  \BibitemOpen
  \bibfield  {author} {\bibinfo {author} {\bibfnamefont {T.~A.}\ \bibnamefont
  {Morgado}}, \bibinfo {author} {\bibfnamefont {T.~G.}\ \bibnamefont
  {Rappoport}}, \bibinfo {author} {\bibfnamefont {S.~S.}\ \bibnamefont
  {Tsirkin}}, \bibinfo {author} {\bibfnamefont {S.}~\bibnamefont {Lannebère}},
  \bibinfo {author} {\bibfnamefont {I.}~\bibnamefont {Souza}}, \ and\ \bibinfo
  {author} {\bibfnamefont {M.~G.}\ \bibnamefont {Silveirinha}},\ }\href@noop {}
  {}\bibinfo {note} {Preprint at \url{http://arXiv.org/abs/2401.13764}
  (2024)}\BibitemShut {NoStop}%
\bibitem [{\citenamefont {Sipe}\ and\ \citenamefont
  {Shkrebtii}(2000)}]{Sipe2000}%
  \BibitemOpen
  \bibfield  {author} {\bibinfo {author} {\bibfnamefont {J.~E.}\ \bibnamefont
  {Sipe}}\ and\ \bibinfo {author} {\bibfnamefont {A.~I.}\ \bibnamefont
  {Shkrebtii}},\ }\href {\doibase 10.1103/PhysRevB.61.5337} {\bibfield
  {journal} {\bibinfo  {journal} {Phys. Rev. B}\ }\textbf {\bibinfo {volume}
  {61}},\ \bibinfo {pages} {5337} (\bibinfo {year} {2000})}\BibitemShut
  {NoStop}%
\bibitem [{\citenamefont {Vasko}\ and\ \citenamefont
  {Raichev}(2006)}]{Vasko2006}%
  \BibitemOpen
  \bibfield  {author} {\bibinfo {author} {\bibfnamefont {F.~T.}\ \bibnamefont
  {Vasko}}\ and\ \bibinfo {author} {\bibfnamefont {O.~E.}\ \bibnamefont
  {Raichev}},\ }\href {\doibase 10.1007/0-387-28041-3} {\emph {\bibinfo {title}
  {Quantum kinetic theory and applications: Electrons, photons, phonons}}}\
  (\bibinfo  {publisher} {Springer Science \& Business Media},\ \bibinfo
  {address} {New York},\ \bibinfo {year} {2006})\BibitemShut {NoStop}%
\bibitem [{\citenamefont {Strikha}\ and\ \citenamefont
  {Vasko}(2010)}]{Strikha2010}%
  \BibitemOpen
  \bibfield  {author} {\bibinfo {author} {\bibfnamefont {M.~V.}\ \bibnamefont
  {Strikha}}\ and\ \bibinfo {author} {\bibfnamefont {F.~T.}\ \bibnamefont
  {Vasko}},\ }\href {\doibase 10.1103/PhysRevB.81.115413} {\bibfield  {journal}
  {\bibinfo  {journal} {Phys. Rev. B}\ }\textbf {\bibinfo {volume} {81}},\
  \bibinfo {pages} {115413} (\bibinfo {year} {2010})}\BibitemShut {NoStop}%
\bibitem [{\citenamefont {Margulis}\ \emph {et~al.}(2017)\citenamefont
  {Margulis}, \citenamefont {Muryumin},\ and\ \citenamefont
  {Gaiduk}}]{Margulis2017}%
  \BibitemOpen
  \bibfield  {author} {\bibinfo {author} {\bibfnamefont {V.~A.}\ \bibnamefont
  {Margulis}}, \bibinfo {author} {\bibfnamefont {E.~E.}\ \bibnamefont
  {Muryumin}}, \ and\ \bibinfo {author} {\bibfnamefont {E.~A.}\ \bibnamefont
  {Gaiduk}},\ }\href {\doibase 10.1088/2040-8986/aa6b6a} {\bibfield  {journal}
  {\bibinfo  {journal} {Journal of Optics}\ }\textbf {\bibinfo {volume} {19}},\
  \bibinfo {pages} {065505} (\bibinfo {year} {2017})}\BibitemShut {NoStop}%
\bibitem [{\citenamefont {Sabbaghi}\ \emph {et~al.}(2018)\citenamefont
  {Sabbaghi}, \citenamefont {Lee},\ and\ \citenamefont
  {Stauber}}]{Sabbaghi2018}%
  \BibitemOpen
  \bibfield  {author} {\bibinfo {author} {\bibfnamefont {M.}~\bibnamefont
  {Sabbaghi}}, \bibinfo {author} {\bibfnamefont {H.-W.}\ \bibnamefont {Lee}}, \
  and\ \bibinfo {author} {\bibfnamefont {T.}~\bibnamefont {Stauber}},\ }\href
  {\doibase 10.1103/PhysRevB.98.075424} {\bibfield  {journal} {\bibinfo
  {journal} {Phys. Rev. B}\ }\textbf {\bibinfo {volume} {98}},\ \bibinfo
  {pages} {075424} (\bibinfo {year} {2018})}\BibitemShut {NoStop}%
\bibitem [{\citenamefont {Sodemann}\ and\ \citenamefont
  {Fu}(2015)}]{Sodemann2015}%
  \BibitemOpen
  \bibfield  {author} {\bibinfo {author} {\bibfnamefont {I.}~\bibnamefont
  {Sodemann}}\ and\ \bibinfo {author} {\bibfnamefont {L.}~\bibnamefont {Fu}},\
  }\href {\doibase 10.1103/PhysRevLett.115.216806} {\bibfield  {journal}
  {\bibinfo  {journal} {Phys. Rev. Lett.}\ }\textbf {\bibinfo {volume} {115}},\
  \bibinfo {pages} {216806} (\bibinfo {year} {2015})}\BibitemShut {NoStop}%
\bibitem [{\citenamefont {K\"onig}\ \emph {et~al.}(2019)\citenamefont
  {K\"onig}, \citenamefont {Dzero}, \citenamefont {Levchenko},\ and\
  \citenamefont {Pesin}}]{Konig2019}%
  \BibitemOpen
  \bibfield  {author} {\bibinfo {author} {\bibfnamefont {E.~J.}\ \bibnamefont
  {K\"onig}}, \bibinfo {author} {\bibfnamefont {M.}~\bibnamefont {Dzero}},
  \bibinfo {author} {\bibfnamefont {A.}~\bibnamefont {Levchenko}}, \ and\
  \bibinfo {author} {\bibfnamefont {D.~A.}\ \bibnamefont {Pesin}},\ }\href
  {\doibase 10.1103/PhysRevB.99.155404} {\bibfield  {journal} {\bibinfo
  {journal} {Phys. Rev. B}\ }\textbf {\bibinfo {volume} {99}},\ \bibinfo
  {pages} {155404} (\bibinfo {year} {2019})}\BibitemShut {NoStop}%
\bibitem [{\citenamefont {Cheng}\ \emph {et~al.}(2019)\citenamefont {Cheng},
  \citenamefont {Wu}, \citenamefont {Zhu},\ and\ \citenamefont
  {Guo}}]{Cheng2019}%
  \BibitemOpen
  \bibfield  {author} {\bibinfo {author} {\bibfnamefont {M.}~\bibnamefont
  {Cheng}}, \bibinfo {author} {\bibfnamefont {S.}~\bibnamefont {Wu}}, \bibinfo
  {author} {\bibfnamefont {Z.-Z.}\ \bibnamefont {Zhu}}, \ and\ \bibinfo
  {author} {\bibfnamefont {G.-Y.}\ \bibnamefont {Guo}},\ }\href {\doibase
  10.1103/PhysRevB.100.035202} {\bibfield  {journal} {\bibinfo  {journal}
  {Phys. Rev. B}\ }\textbf {\bibinfo {volume} {100}},\ \bibinfo {pages}
  {035202} (\bibinfo {year} {2019})}\BibitemShut {NoStop}%
\bibitem [{\citenamefont {Li}\ \emph {et~al.}(2022)\citenamefont {Li},
  \citenamefont {Gao}, \citenamefont {Gu}, \citenamefont {Zhang}, \citenamefont
  {Iitaka},\ and\ \citenamefont {Liu}}]{Li2022}%
  \BibitemOpen
  \bibfield  {author} {\bibinfo {author} {\bibfnamefont {Z.}~\bibnamefont
  {Li}}, \bibinfo {author} {\bibfnamefont {Y.}~\bibnamefont {Gao}}, \bibinfo
  {author} {\bibfnamefont {Y.}~\bibnamefont {Gu}}, \bibinfo {author}
  {\bibfnamefont {S.}~\bibnamefont {Zhang}}, \bibinfo {author} {\bibfnamefont
  {T.}~\bibnamefont {Iitaka}}, \ and\ \bibinfo {author} {\bibfnamefont {W.~M.}\
  \bibnamefont {Liu}},\ }\href {\doibase 10.1103/PhysRevB.105.125201}
  {\bibfield  {journal} {\bibinfo  {journal} {Phys. Rev. B}\ }\textbf {\bibinfo
  {volume} {105}},\ \bibinfo {pages} {125201} (\bibinfo {year}
  {2022})}\BibitemShut {NoStop}%
\bibitem [{\citenamefont {Pantale\'on}\ \emph {et~al.}(2021)\citenamefont
  {Pantale\'on}, \citenamefont {Low},\ and\ \citenamefont
  {Guinea}}]{Pantaleon2021}%
  \BibitemOpen
  \bibfield  {author} {\bibinfo {author} {\bibfnamefont {P.~A.}\ \bibnamefont
  {Pantale\'on}}, \bibinfo {author} {\bibfnamefont {T.}~\bibnamefont {Low}}, \
  and\ \bibinfo {author} {\bibfnamefont {F.}~\bibnamefont {Guinea}},\ }\href
  {\doibase 10.1103/PhysRevB.103.205403} {\bibfield  {journal} {\bibinfo
  {journal} {Phys. Rev. B}\ }\textbf {\bibinfo {volume} {103}},\ \bibinfo
  {pages} {205403} (\bibinfo {year} {2021})}\BibitemShut {NoStop}%
\bibitem [{\citenamefont {He}\ and\ \citenamefont {Weng}(2021)}]{He2021}%
  \BibitemOpen
  \bibfield  {author} {\bibinfo {author} {\bibfnamefont {Z.}~\bibnamefont
  {He}}\ and\ \bibinfo {author} {\bibfnamefont {H.}~\bibnamefont {Weng}},\
  }\href {\doibase 10.1038/s41535-021-00403-9} {\bibfield  {journal} {\bibinfo
  {journal} {npj Quantum Materials}\ }\textbf {\bibinfo {volume} {6}},\
  \bibinfo {pages} {101} (\bibinfo {year} {2021})}\BibitemShut {NoStop}%
\bibitem [{\citenamefont {Almasov}\ and\ \citenamefont
  {Dykman}(1971)}]{Almasov1971}%
  \BibitemOpen
  \bibfield  {author} {\bibinfo {author} {\bibfnamefont {L.~A.}\ \bibnamefont
  {Almasov}}\ and\ \bibinfo {author} {\bibfnamefont {I.~M.}\ \bibnamefont
  {Dykman}},\ }\href {\doibase 10.1002/pssb.2220480213} {\bibfield  {journal}
  {\bibinfo  {journal} {Phys. Status Solidi B}\ }\textbf {\bibinfo {volume}
  {48}},\ \bibinfo {pages} {563} (\bibinfo {year} {1971})}\BibitemShut
  {NoStop}%
\bibitem [{\citenamefont {Isobe}\ \emph {et~al.}(2020)\citenamefont {Isobe},
  \citenamefont {Xu},\ and\ \citenamefont {Fu}}]{Liang2020}%
  \BibitemOpen
  \bibfield  {author} {\bibinfo {author} {\bibfnamefont {H.}~\bibnamefont
  {Isobe}}, \bibinfo {author} {\bibfnamefont {S.-Y.}\ \bibnamefont {Xu}}, \
  and\ \bibinfo {author} {\bibfnamefont {L.}~\bibnamefont {Fu}},\ }\href
  {\doibase 10.1126/sciadv.aay2497} {\bibfield  {journal} {\bibinfo  {journal}
  {Science Advances}\ }\textbf {\bibinfo {volume} {6}},\ \bibinfo {pages}
  {aay2497} (\bibinfo {year} {2020})}\BibitemShut {NoStop}%
\bibitem [{See()}]{SeeSI}%
  \BibitemOpen
  \href@noop {} {}\bibinfo {note} {See Supplemental Material for additional
  theoretical details as well as
  Refs.~\cite{Ashcroft,Sinitsyn2007,Nagaosa2010}.}\BibitemShut {Stop}%
\bibitem [{\citenamefont {Ma}\ \emph {et~al.}(2018)\citenamefont {Ma},
  \citenamefont {Xu}, \citenamefont {Shen}, \citenamefont {MacNeill},
  \citenamefont {Fatemi}, \citenamefont {Chang}, \citenamefont {Valdivia},
  \citenamefont {Wu}, \citenamefont {Du}, \citenamefont {Hsu}, \citenamefont
  {Fang}, \citenamefont {Gibson}, \citenamefont {Watanabe}, \citenamefont
  {Taniguchi}, \citenamefont {Cava}, \citenamefont {Kaxiras}, \citenamefont
  {Lu}, \citenamefont {Lin}, \citenamefont {Fu}, \citenamefont {Gedik},\ and\
  \citenamefont {Jarillo-Herrero}}]{QiongMa2018}%
  \BibitemOpen
  \bibfield  {author} {\bibinfo {author} {\bibfnamefont {Q.}~\bibnamefont
  {Ma}}, \bibinfo {author} {\bibfnamefont {S.-Y.}\ \bibnamefont {Xu}}, \bibinfo
  {author} {\bibfnamefont {H.}~\bibnamefont {Shen}}, \bibinfo {author}
  {\bibfnamefont {D.}~\bibnamefont {MacNeill}}, \bibinfo {author}
  {\bibfnamefont {V.}~\bibnamefont {Fatemi}}, \bibinfo {author} {\bibfnamefont
  {T.-R.}\ \bibnamefont {Chang}}, \bibinfo {author} {\bibfnamefont {A.~M.~M.}\
  \bibnamefont {Valdivia}}, \bibinfo {author} {\bibfnamefont {S.}~\bibnamefont
  {Wu}}, \bibinfo {author} {\bibfnamefont {Z.}~\bibnamefont {Du}}, \bibinfo
  {author} {\bibfnamefont {C.-H.}\ \bibnamefont {Hsu}}, \bibinfo {author}
  {\bibfnamefont {S.}~\bibnamefont {Fang}}, \bibinfo {author} {\bibfnamefont
  {Q.~D.}\ \bibnamefont {Gibson}}, \bibinfo {author} {\bibfnamefont
  {K.}~\bibnamefont {Watanabe}}, \bibinfo {author} {\bibfnamefont
  {T.}~\bibnamefont {Taniguchi}}, \bibinfo {author} {\bibfnamefont {R.~J.}\
  \bibnamefont {Cava}}, \bibinfo {author} {\bibfnamefont {E.}~\bibnamefont
  {Kaxiras}}, \bibinfo {author} {\bibfnamefont {H.-Z.}\ \bibnamefont {Lu}},
  \bibinfo {author} {\bibfnamefont {H.}~\bibnamefont {Lin}}, \bibinfo {author}
  {\bibfnamefont {L.}~\bibnamefont {Fu}}, \bibinfo {author} {\bibfnamefont
  {N.}~\bibnamefont {Gedik}}, \ and\ \bibinfo {author} {\bibfnamefont
  {P.}~\bibnamefont {Jarillo-Herrero}},\ }\href {\doibase
  10.1038/s41586-018-0807-6} {\bibfield  {journal} {\bibinfo  {journal}
  {Nature}\ }\textbf {\bibinfo {volume} {565}},\ \bibinfo {pages} {337}
  (\bibinfo {year} {2018})}\BibitemShut {NoStop}%
\bibitem [{\citenamefont {Kang}\ \emph {et~al.}(2019)\citenamefont {Kang},
  \citenamefont {Li}, \citenamefont {Sohn}, \citenamefont {Shan},\ and\
  \citenamefont {Mak}}]{MakKinFai2019}%
  \BibitemOpen
  \bibfield  {author} {\bibinfo {author} {\bibfnamefont {K.}~\bibnamefont
  {Kang}}, \bibinfo {author} {\bibfnamefont {T.}~\bibnamefont {Li}}, \bibinfo
  {author} {\bibfnamefont {E.}~\bibnamefont {Sohn}}, \bibinfo {author}
  {\bibfnamefont {J.}~\bibnamefont {Shan}}, \ and\ \bibinfo {author}
  {\bibfnamefont {K.~F.}\ \bibnamefont {Mak}},\ }\href {\doibase
  10.1038/s41563-019-0294-7} {\bibfield  {journal} {\bibinfo  {journal} {Nature
  Materials}\ }\textbf {\bibinfo {volume} {18}},\ \bibinfo {pages} {324}
  (\bibinfo {year} {2019})}\BibitemShut {NoStop}%
\bibitem [{\citenamefont {Sinitsyn}\ \emph {et~al.}(2006)\citenamefont
  {Sinitsyn}, \citenamefont {Niu},\ and\ \citenamefont
  {MacDonald}}]{Sinitsyn2006}%
  \BibitemOpen
  \bibfield  {author} {\bibinfo {author} {\bibfnamefont {N.~A.}\ \bibnamefont
  {Sinitsyn}}, \bibinfo {author} {\bibfnamefont {Q.}~\bibnamefont {Niu}}, \
  and\ \bibinfo {author} {\bibfnamefont {A.~H.}\ \bibnamefont {MacDonald}},\
  }\href {\doibase 10.1103/PhysRevB.73.075318} {\bibfield  {journal} {\bibinfo
  {journal} {Phys. Rev. B}\ }\textbf {\bibinfo {volume} {73}},\ \bibinfo
  {pages} {075318} (\bibinfo {year} {2006})}\BibitemShut {NoStop}%
\bibitem [{\citenamefont {Sprott}(1997)}]{Sprott1997}%
  \BibitemOpen
  \bibfield  {author} {\bibinfo {author} {\bibfnamefont {J.~C.}\ \bibnamefont
  {Sprott}},\ }\href {\doibase 10.1119/1.18585} {\bibfield  {journal} {\bibinfo
   {journal} {American Journal of Physics}\ }\textbf {\bibinfo {volume} {65}},\
  \bibinfo {pages} {537} (\bibinfo {year} {1997})}\BibitemShut {NoStop}%
\bibitem [{\citenamefont {Gao}\ \emph {et~al.}(2014)\citenamefont {Gao},
  \citenamefont {Yang},\ and\ \citenamefont {Niu}}]{Gao2014}%
  \BibitemOpen
  \bibfield  {author} {\bibinfo {author} {\bibfnamefont {Y.}~\bibnamefont
  {Gao}}, \bibinfo {author} {\bibfnamefont {S.~A.}\ \bibnamefont {Yang}}, \
  and\ \bibinfo {author} {\bibfnamefont {Q.}~\bibnamefont {Niu}},\ }\href
  {\doibase 10.1103/PhysRevLett.112.166601} {\bibfield  {journal} {\bibinfo
  {journal} {Phys. Rev. Lett.}\ }\textbf {\bibinfo {volume} {112}},\ \bibinfo
  {pages} {166601} (\bibinfo {year} {2014})}\BibitemShut {NoStop}%
\bibitem [{\citenamefont {Ashcroft}\ and\ \citenamefont
  {Mermin}(1976)}]{Ashcroft}%
  \BibitemOpen
  \bibfield  {author} {\bibinfo {author} {\bibfnamefont {N.~W.}\ \bibnamefont
  {Ashcroft}}\ and\ \bibinfo {author} {\bibfnamefont {N.~D.}\ \bibnamefont
  {Mermin}},\ }\href@noop {} {\emph {\bibinfo {title} {Solid state physics}}}\
  (\bibinfo  {publisher} {Holt, Rinehart and Winston},\ \bibinfo {address} {New
  York},\ \bibinfo {year} {1976})\BibitemShut {NoStop}%
\bibitem [{\citenamefont {Sinitsyn}(2007)}]{Sinitsyn2007}%
  \BibitemOpen
  \bibfield  {author} {\bibinfo {author} {\bibfnamefont {N.~A.}\ \bibnamefont
  {Sinitsyn}},\ }\href {\doibase 10.1088/0953-8984/20/02/023201} {\bibfield
  {journal} {\bibinfo  {journal} {Journal of Physics: Condensed Matter}\
  }\textbf {\bibinfo {volume} {20}},\ \bibinfo {pages} {023201} (\bibinfo
  {year} {2007})}\BibitemShut {NoStop}%
\bibitem [{\citenamefont {Nagaosa}\ \emph {et~al.}(2010)\citenamefont
  {Nagaosa}, \citenamefont {Sinova}, \citenamefont {Onoda}, \citenamefont
  {MacDonald},\ and\ \citenamefont {Ong}}]{Nagaosa2010}%
  \BibitemOpen
  \bibfield  {author} {\bibinfo {author} {\bibfnamefont {N.}~\bibnamefont
  {Nagaosa}}, \bibinfo {author} {\bibfnamefont {J.}~\bibnamefont {Sinova}},
  \bibinfo {author} {\bibfnamefont {S.}~\bibnamefont {Onoda}}, \bibinfo
  {author} {\bibfnamefont {A.~H.}\ \bibnamefont {MacDonald}}, \ and\ \bibinfo
  {author} {\bibfnamefont {N.~P.}\ \bibnamefont {Ong}},\ }\href {\doibase
  10.1103/RevModPhys.82.1539} {\bibfield  {journal} {\bibinfo  {journal} {Rev.
  Mod. Phys.}\ }\textbf {\bibinfo {volume} {82}},\ \bibinfo {pages} {1539}
  (\bibinfo {year} {2010})}\BibitemShut {NoStop}%
\end{thebibliography}%

\onecolumngrid
\newpage
\pagebreak
\widetext
\begin{center}
\textbf{\large Supplemental Information for ``Skew-scattering Pockels effect and metallic electro-optics in gapped bilayer graphene''}
\end{center}
\setcounter{equation}{0}
\setcounter{table}{0}
\setcounter{figure}{0}
\makeatletter 
\renewcommand{\thefigure}{S\arabic{figure}}
\renewcommand{\theequation}{S\arabic{equation}}
\renewcommand{\thetable}{S\arabic{table}}
\renewcommand{\bibnumfmt}[1]{[S#1]}
\renewcommand{\theHequation}{Supplement.\theequation}
\renewcommand{\theHfigure}{Supplement.\thefigure}
\renewcommand{\theHtable}{Supplement.\thetable}


\subsection{The electro-optic effect and the Boltzmann equation}
Here we review the standard derivation of nonlinear responses up to third order with the Boltzmann equation, including both the symmetric scattering and the antisymmetric skew scattering in order to get the coefficients for the electro-optic effect. We begin with the Boltzmann equation for a spatially homogeneous system~\cite{Ashcroft,Sinitsyn2007,Nagaosa2010},
\begin{equation}
\label{eq:S-Boltzmann}
\frac{\partial }{\partial t} f \left({\bf k},t \right) - \frac{e}{\hbar} \boldsymbol{\mathcal{E}} \left(t \right) \cdot \partial_{\bf k} f \left({\bf k},t \right) = \mathcal{I}\{ f \left({\bf k},t \right) \},
\end{equation} 
where $\boldsymbol{\mathcal{E}} (t) = \boldsymbol{\mathcal{E}}_{\rm AC} (t) + \vec E$ is the total electric field including an oscillating part $\boldsymbol{\mathcal{E}}_{AC}(t) = \boldsymbol{\mathcal{E}}^{(0)} e^{i\omega t} + c.c.$ and an applied DC field $\vec E$, and $\mathcal{I}\{ f \left({\bf k},t \right) \}$ is the collision integral~\cite{Ashcroft,Sinitsyn2007,Nagaosa2010},
\begin{equation}
\mathcal{I}\{ f \left({\bf k},t \right) \} = - \sum_{{\bf k}'}\left\{ \left( w^{\mathrm{S}}_{{\bf k}^{\prime}{\bf k} } + w^{\mathrm{A}}_{{\bf k}^{\prime}{\bf k} }\right) f \left({\bf k} ,t \right) - \left( w^{\mathrm{S}}_{{\bf k}{\bf k}^{\prime} } + w^{\mathrm{A}}_{{\bf k}{\bf k}^{\prime} }\right) f \left({\bf k}^{\prime} ,t \right) \right\},
\end{equation}
with $w^{\mathrm{S}}_{{\bf k}{\bf k}^{\prime} }$ and $w^{\mathrm{A}}_{{\bf k}{\bf k}^{\prime} }$ for symmetric and anti-symmetric scattering rate, respectively. The collision integral above will be simplified with conservation of probabilities $\sum_{{\bf k}'} w^{\mathrm{A}}_{{\bf k}^{\prime}{\bf k} } = 0$~\cite{Konig2019,Liang2020} in the following.

To study both the Pockels and the Kerr effect, we need to solve the equation up to the third order of electric field. We solve Eq.~(\ref{eq:S-Boltzmann}) perturbatively with the following ansatz~\cite{Liang2020}
\begin{equation}
f \left({\bf k},t \right) =  f_{0} \left({\bf k} \right) + \sum_{\ell, m} f_{\ell}^{(m)} \left({\bf k},t \right), \quad  f_{\ell}^{(m)} \left({\bf k},t \right) = \sum_{\nu} f_{\nu, \ell}^{(m)} \left({\bf k},t \right),
\end{equation}
where $f_{0} \left({\bf k} \right)$ is the distribution function at equilibrium, $\nu$ is the frequency of the corresponding component of the distribution function, $\ell$ is the order of the electric field, and $m$ is the order of anti-symmetric scattering $w^{\mathrm{A}}_{{\bf k}{\bf k}^{\prime} }$. For example,  $f_{0, 1}^{(0)} \left({\bf k},t \right)$ is the purely symmetrically scattered distribution function up to first order of electric field of $0$ frequency (DC component).

In the perturbative fashion, we can decompose the Boltzmann equation in Eq.~(\ref{eq:S-Boltzmann}) in orders of $\nu$, $\ell$ and $m$. We first try to solve the equation at the first order of ${\bf E}$, i.e., $\ell = 1$. The symmetric scattering part ($m=0$) reads
\begin{equation}
\label{eq:BE-f1(0)}
\frac{\partial }{\partial t} f_{1}^{(0)} \left({\bf k},t \right) - \frac{e}{\hbar} \boldsymbol{\mathcal{E}} \left(t \right) \cdot \partial_{\bf k} f_{0} \left({\bf k} \right) =  - \sum_{{\bf k}'} w^{\mathrm{S}}_{{\bf k}^{\prime}{\bf k} } \left( f_{1}^{(0)} \left({\bf k} ,t \right) - f_{1}^{(0)}\left({\bf k}^{\prime} ,t \right) \right)  ,
\end{equation} 
while the anti-symmetric (skew) scattering part follows
\begin{equation}
\label{eq:BE-f1(1)}
\frac{\partial }{\partial t} f_{1}^{(1)}\left({\bf k},t \right) =  - \sum_{{\bf k}'}\left\{ w^{\mathrm{S}}_{{\bf k}^{\prime}{\bf k} } \left( f_{1}^{(1)} \left({\bf k} ,t \right) - f_{1}^{(1)}  \left({\bf k}^{\prime} ,t \right) \right) - w^{\mathrm{A}}_{{\bf k}{\bf k}^{\prime} } f_{1}^{(0)} \left({\bf k}^{\prime} ,t \right) \right\}.
\end{equation} 
Components of different frequencies can be separated in a straightforward way. For example, we can break Eq.~(\ref{eq:BE-f1(0)}) into the following equations,
\begin{align}
\frac{\partial }{\partial t} f_{\omega, 1}^{(0)} \left({\bf k},t \right) - \frac{e}{\hbar} \boldsymbol{\mathcal{E}}^{(0)} e^{i \omega t} \cdot \partial_{\bf k} f_{0} \left({\bf k} \right) & =  - \sum_{{\bf k}'} w^{\mathrm{S}}_{{\bf k}^{\prime}{\bf k} } \left( f_{\omega, 1}^{(0)} \left({\bf k} ,t \right) - f_{\omega, 1}^{(0)}\left({\bf k}^{\prime} ,t \right) \right)  , \\
\frac{\partial }{\partial t} f_{\mathrm{0}, 1}^{(0)} \left({\bf k},t \right) - \frac{e}{\hbar} {\bf E} \cdot \partial_{\bf k} f_{0} \left({\bf k} \right) & =  - \sum_{{\bf k}'} w^{\mathrm{S}}_{{\bf k}^{\prime}{\bf k} } \left( f_{\mathrm{0}, 1}^{(0)} \left({\bf k} ,t \right) - f_{\mathrm{0}, 1}^{(0)}\left({\bf k}^{\prime} ,t \right) \right)  , \\ 
\frac{\partial }{\partial t} f_{-\omega, 1}^{(0)} \left({\bf k},t \right) - \frac{e}{\hbar} {\boldsymbol{\mathcal{E}}^{(0)}}^{*} e^{ - i \omega t} \cdot \partial_{\bf k} f_{0} \left({\bf k} \right) & =  - \sum_{{\bf k}'} w^{\mathrm{S}}_{{\bf k}^{\prime}{\bf k} } \left( f_{-\omega, 1}^{(0)} \left({\bf k} ,t \right) - f_{-\omega, 1}^{(0)}\left({\bf k}^{\prime} ,t \right) \right)  .  
\end{align}

To solve for the distribution functions, we take the relaxation time approximation~\cite{Ashcroft,Sinitsyn2007,Nagaosa2010,Konig2019,Liang2020},
\begin{equation}
\label{eq:RTA}
\sum_{{\bf k}'} w^{\mathrm{S}}_{{\bf k}^{\prime}{\bf k} } \left( f_{\nu, \ell}^{(m)} \left({\bf k} ,t \right) - f _{\nu, \ell}^{(m)}\left({\bf k}^{\prime} ,t \right) \right) =  \frac{1}{ \tau } f_{\nu, \ell}^{(m)} \left({\bf k} ,t \right) , \quad  \frac{1}{\tau} = \langle \sum_{{\bf k}'} w^{\mathrm{S}}_{{\bf k}^{\prime}{\bf k} } \left( 1 - \cos \theta_{{\bf v} {\bf v}^{\prime}}\right) \rangle ,
\end{equation}
where $\theta_{{\bf v} {\bf v}^{\prime}}$ is the angle between ${\bf v} \left( {\bf k} \right)$ and ${\bf v}  \left( {\bf k^{\prime} } \right)$, and $\langle ... \rangle$ indicates taking average on the energy contour. We thereby take the single relaxation time approximation where all components of the distribution function share a single $\tau$ and the constant relaxation time approximation in which $\tau$ is momentum-independent~\cite{Konig2019}. In the following, we use the short-hand notation $\tau_{\omega} = \frac{\tau}{1 + i \omega \tau}$.

With the relaxation time approximation introduced above, we have 
\begin{equation}
f_{\omega, 1}^{(0)} \left({\bf k},t \right) =  \frac{e}{\hbar}  e^{i \omega t} \tau_{\omega} {\boldsymbol{\mathcal{E}}^{(0)}} \cdot \partial_{\bf k} f_{0} \left({\bf k} \right) , \quad f_{0, 1}^{(0)} \left({\bf k},t \right) =  \frac{e}{\hbar}  \tau {\bf E} \cdot \partial_{\bf k} f_{0} \left({\bf k} \right) , \quad f_{-\omega, 1}^{(0)} \left({\bf k},t \right) = \left( f_{\omega, 1}^{(0)} \left({\bf k},t \right) \right)^{*}.
\end{equation}

Similarly, for $\ell = 2$, we have 
\begin{align}
\frac{\partial }{\partial t} f_{2}^{(0)} \left({\bf k},t \right) - \frac{e}{\hbar} \boldsymbol{\mathcal{E}} \left(t \right) \cdot \partial_{\bf k} f_{1}^{(0)} \left({\bf k,t} \right) &=  - \sum_{{\bf k}'} w^{\mathrm{S}}_{{\bf k}^{\prime}{\bf k} } \left( f_{2}^{(0)} \left({\bf k} ,t \right) - f_{2}^{(0)}\left({\bf k}^{\prime} ,t \right) \right)  , \\
\frac{\partial }{\partial t} f_{2}^{(1)}\left({\bf k},t \right) - \frac{e}{\hbar} \boldsymbol{\mathcal{E}} \left(t \right) \cdot \partial_{\bf k} f_{1}^{(1)} \left({\bf k,t} \right)  & =  - \sum_{{\bf k}'}\left\{ w^{\mathrm{S}}_{{\bf k}^{\prime}{\bf k} } \left( f_{2}^{(1)} \left({\bf k} ,t \right) - f_{2}^{(1)}  \left({\bf k}^{\prime} ,t \right) \right) - w^{\mathrm{A}}_{{\bf k}{\bf k}^{\prime} } f_{2}^{(0)} \left({\bf k}^{\prime} ,t \right) \right\}.
\end{align} 
Following the same procedure, break up the equation for different frequencies and take the relaxation time approximation, for symmetric scattering we can get
\begin{align}
f_{2\omega, 2}^{(0)} \left({\bf k},t \right) &= \frac{e^{2} }{\hbar^{2} }e^{2 i \omega t} \tau_{2 \omega} {\boldsymbol{\mathcal{E}}^{(0)}} \cdot \partial_{\bf k} \tau_{\omega} {\boldsymbol{\mathcal{E}}^{(0)}} \cdot \partial_{\bf k} f_{0} \left({\bf k} \right) , \\
f_{\omega, 2}^{(0)} \left({\bf k},t \right) &= \frac{e^{2} }{\hbar^{2} } e^{i \omega t} \tau_{\omega} \bigg\{ {\boldsymbol{\mathcal{E}}^{(0)}} \cdot \partial_{\bf k} \tau {\bf E} \cdot \partial_{\bf k} f_{0} \left({\bf k} \right) + {\bf E} \cdot \partial_{\bf k} \tau_{\omega} {\boldsymbol{\mathcal{E}}^{(0)}} \cdot \partial_{\bf k} f_{0} \left({\bf k} \right)   \bigg\}  , \\
f_{0, 2}^{(0)} \left({\bf k},t \right) &= \frac{e^{2} }{\hbar^{2} } \tau \bigg\{ {\boldsymbol{\mathcal{E}}^{(0)}} \cdot \partial_{\bf k} \tau_{-\omega} {\boldsymbol{\mathcal{E}}^{(0)}}^{*} \cdot \partial_{\bf k}f_{0} \left({\bf k} \right) + {\bf E} \cdot \partial_{\bf k} \tau {\bf E} \cdot \partial_{\bf k} f_{0} \left({\bf k} \right) + {\boldsymbol{\mathcal{E}}^{(0)}}^{*} \cdot \partial_{\bf k} \tau_{\omega} {\boldsymbol{\mathcal{E}}^{(0)}} \cdot \partial_{\bf k} f_{0} \left({\bf k} \right)  \bigg\}.
\end{align}
and for antisymmetric skew scattering we have
\begin{align}
f_{2\omega, 2}^{(1)} \left({\bf k},t \right) &= \frac{e^{2} }{\hbar^{2} }e^{2 i \omega t} \tau_{2 \omega} \bigg\{ {\boldsymbol{\mathcal{E}}^{(0)}} \cdot \partial_{\bf k} \tau_{ \omega} \sum_{{\bf k}'} w^{\mathrm{A}}_{{\bf k}{\bf k}^{\prime} } \tau_{ \omega} {\boldsymbol{\mathcal{E}}^{(0)}} \cdot \partial_{{\bf k}^{\prime}} f_{0} \left({\bf k}^{\prime} \right) + \sum_{{\bf k}'} w^{\mathrm{A}}_{{\bf k}{\bf k}^{\prime} } \tau_{2 \omega} {\boldsymbol{\mathcal{E}}^{(0)}} \cdot \partial_{{\bf k}^{\prime}} \tau_{ \omega} {\boldsymbol{\mathcal{E}}^{(0)}} \cdot \partial_{{\bf k}^{\prime}} f_{0} \left({\bf k}^{\prime} \right) \bigg\} , \\
f_{\omega, 2}^{(1)} \left({\bf k},t \right) &= \frac{e^{2} }{\hbar^{2} }e^{i \omega t} \tau_{ \omega} \bigg\{ {\boldsymbol{\mathcal{E}}^{(0)}} \cdot \partial_{\bf k} \tau \sum_{{\bf k}'} w^{\mathrm{A}}_{{\bf k}{\bf k}^{\prime} } \tau {\bf E} \cdot \partial_{{\bf k}^{\prime}} f_{0} \left({\bf k}^{\prime} \right) + {\bf E} \cdot \partial_{\bf k} \tau_{ \omega} \sum_{{\bf k}'} w^{\mathrm{A}}_{{\bf k}{\bf k}^{\prime} } \tau_{ \omega} {\boldsymbol{\mathcal{E}}^{(0)}} \cdot \partial_{{\bf k}^{\prime}} f_{0} \left({\bf k}^{\prime} \right)  \nonumber \\
& +  \sum_{{\bf k}'} w^{\mathrm{A}}_{{\bf k}{\bf k}^{\prime} } \tau_{ \omega} \big[ {\boldsymbol{\mathcal{E}}^{(0)}} \cdot \partial_{{\bf k}^{\prime}} \tau {\bf E} \cdot \partial_{{\bf k}^{\prime}} f_{0} \left({\bf k}^{\prime} \right) + {\bf E} \cdot \partial_{{\bf k}^{\prime}} \tau_{ \omega} {\boldsymbol{\mathcal{E}}^{(0)}} \cdot \partial_{{\bf k}^{\prime}}  f_{0} \left({\bf k}^{\prime} \right) \big] \bigg\} ,  \\
f_{0, 2}^{(1)} \left({\bf k},t \right) &= \frac{e^{2} }{\hbar^{2} } \tau \bigg\{ {\boldsymbol{\mathcal{E}}^{(0)}} \cdot \partial_{\bf k} \tau_{- \omega} \sum_{{\bf k}'} w^{\mathrm{A}}_{{\bf k}{\bf k}^{\prime} } \tau_{- \omega} {\boldsymbol{\mathcal{E}}^{(0)}}^{*} \cdot \partial_{\bf k^{\prime}}  f_{0} \left({\bf k}^{\prime} \right) + {\bf E} \cdot \partial_{\bf k} \tau \sum_{{\bf k}'} w^{\mathrm{A}}_{{\bf k}{\bf k}^{\prime} } \tau {\bf E} \cdot \partial_{{\bf k}^{\prime}} f_{0} \left({\bf k}^{\prime} \right) \nonumber \\
& + {\boldsymbol{\mathcal{E}}^{(0)}}^{*} \cdot \partial_{\bf k} \tau_{ \omega} \sum_{{\bf k}'} w^{\mathrm{A}}_{{\bf k}{\bf k}^{\prime} } \tau_{ \omega} {\boldsymbol{\mathcal{E}}^{(0)}} \cdot \partial_{{\bf k}^{\prime}} f_{0} \left({\bf k}^{\prime} \right) + \sum_{{\bf k}'} w^{\mathrm{A}}_{{\bf k}{\bf k}^{\prime} } \tau {\boldsymbol{\mathcal{E}}^{(0)}} \cdot \partial_{{\bf k}^{\prime}} \tau_{- \omega} {\boldsymbol{\mathcal{E}}^{(0)}}^{*} \cdot \partial_{{\bf k}^{\prime}}  f_{0} \left({\bf k}^{\prime} \right) \nonumber \\
& + \sum_{{\bf k}'} w^{\mathrm{A}}_{{\bf k}{\bf k}^{\prime} } \tau {\bf E} \cdot \partial_{{\bf k}^{\prime}} \tau {\bf E} \cdot \partial_{{\bf k}^{\prime}}  f_{0} \left({\bf k}^{\prime} \right) + \sum_{{\bf k}'} w^{\mathrm{A}}_{{\bf k}{\bf k}^{\prime} } \tau {\boldsymbol{\mathcal{E}}^{(0)}}^{*} \cdot \partial_{{\bf k}^{\prime}} \tau_{ \omega} {\boldsymbol{\mathcal{E}}^{(0)}} \cdot \partial_{{\bf k}^{\prime}}  f_{0} \left({\bf k}^{\prime} \right) \bigg\}.
\end{align}

The above distribution functions (for $\ell = 1$ and $\ell = 2$) contribute to the Pockels effect. We now examine distribution functions for $\ell = 3$. These will contribute to the Kerr effect discussed in the main text. In so doing, we will focus on symmetric scattering $m = 0$, as the antisymmetric skew scattering contribution is disallowed by time-reversal symmetry,
\begin{equation}
\label{eq:BE-f3(0)}
\frac{\partial }{\partial t} f_{3}^{(0)} \left({\bf k},t \right) - \frac{e}{\hbar} \boldsymbol{\mathcal{E}} \left(t \right) \cdot \partial_{\bf k} f_{2}^{(0)} \left({\bf k,t} \right) =  - \sum_{{\bf k}'} w^{\mathrm{S}}_{{\bf k}^{\prime}{\bf k} } \left( f_{3}^{(0)} \left({\bf k} ,t \right) - f_{3}^{(0)}\left({\bf k}^{\prime} ,t \right) \right) ,
\end{equation} 
and the distribution function is solved in the same way,
\begin{align}
\label{eq:f-omega-3-0}
f_{\omega, 3}^{(0)} \left({\bf k},t \right) =& \frac{e^{3} }{\hbar^{3} } e^{i \omega t} \tau_{ \omega} \bigg\{ {\boldsymbol{\mathcal{E}}^{(0)}} \cdot \partial_{\bf k}  \tau \bigg( {\boldsymbol{\mathcal{E}}^{(0)}} \cdot \partial_{\bf k} \tau_{ - \omega} {\boldsymbol{\mathcal{E}}^{(0)}}^{*} \cdot \partial_{\bf k}f_{0} \left({\bf k} \right) + {\bf E} \cdot \partial_{\bf k} \tau {\bf E} \cdot \partial_{\bf k} f_{0} \left({\bf k} \right) + {\boldsymbol{\mathcal{E}}^{(0)}}^{*} \cdot \partial_{\bf k} \tau_{ \omega} {\boldsymbol{\mathcal{E}}^{(0)}} \cdot \partial_{\bf k} f_{0} \left({\bf k} \right)  \bigg) \nonumber \\
+ & {\bf E} \cdot \partial_{\bf k} \tau_{ \omega} \left( {\boldsymbol{\mathcal{E}}^{(0)}} \cdot \partial_{\bf k} \tau {\bf E} \cdot \partial_{\bf k} f_{0} \left({\bf k} \right) + {\bf E} \cdot \partial_{\bf k}\tau_{ \omega} {\boldsymbol{\mathcal{E}}^{(0)}} \cdot \partial_{\bf k} f_{0} \left({\bf k} \right)   \right) + {\boldsymbol{\mathcal{E}}^{(0)}}^{*} \cdot \partial_{\bf k} \tau_{ 2 \omega} {\boldsymbol{\mathcal{E}}^{(0)}} \cdot \partial_{\bf k} \tau_{ \omega} {\boldsymbol{\mathcal{E}}^{(0)}} \cdot \partial_{\bf k} f_{0} \left({\bf k} \right) \bigg\}.
\end{align}

For the electro-optic effect, we focus on the electric current of frequency $\omega$, ${\bf j} \left( t \right) = {\bf j}^{\omega} e^{i \omega t} + \mathrm{c.c.}$, where ${j}^{\omega}_{\alpha } = \sigma_{\alpha\beta} {\mathcal{E}}^{(0)}_{\beta}$. The conductivity $\sigma_{\alpha\beta}$ is the sum of a field-free part $\sigma^{(0)} \left( \omega \right)$ and the electro-optic contribution $\delta \sigma_{\alpha \beta}$, $\sigma_{\alpha\beta} \left( \omega \right) = \sigma^{(0)}_{\alpha\beta}  \left( \omega \right) + \delta\sigma_{\alpha\beta}  \left( \omega \right)$. $\delta \sigma_{\alpha \beta}$ can be expressed with the Pockels and the Kerr coefficients $P_{\alpha\beta\gamma}$ and $K_{\alpha\beta\gamma\delta}$, 
\begin{equation}
\delta\sigma_{\alpha\beta}  \left( \omega \right) = P_{\alpha\beta\gamma} E_{\gamma}+ K_{\alpha\beta\gamma\delta} E_{\gamma} E_{\delta} + \mathcal{O}(E^3) .
\end{equation}
In order to extract the coefficients $P_{\alpha\beta\gamma}$ and $K_{\alpha\beta\gamma\delta}$, we evaluate the current in the semiclassical picture,
\begin{equation}
\label{eq:S-j(t)}
{\bf j} (t) = - e \sum_{{\bf k}} \bigg\{ \frac{1}{\hbar} \nabla_{ {\bf k} } \varepsilon + \frac{e}{\hbar} \boldsymbol{\mathcal{E}}^{(0)} \left(t \right) \times {\bf \Omega} \bigg\}  f ({\bf k}, t),
\end{equation}
where ${\bf \Omega}$ is Berry curvature. We assume the oscillating part of the electric field is much weaker than the DC field, $\vert {\boldsymbol{\mathcal{E}}^{(0)}} \vert \ll \vert {\bf E} \vert$, so that the AC Kerr effect can be ignored. Therefore, third-order terms proportional to ${\mathcal{E}}^{(0)} {\mathcal{E}^{(0)}}^{*} {\mathcal{E}}^{(0)}$ are dropped.  

First, the bare conductivity $\sigma^{(0)}_{\alpha\beta}$ can be obtained from the field-free current ${\bf j}_{\omega, 1} (t) = - e \sum_{{\bf k}} \frac{1}{\hbar} \nabla_{ {\bf k} } \varepsilon f_{\omega, 1}^{(0)} ({\bf k}, t),$
\begin{equation}
\sigma^{(0)}_{\alpha\beta} = - \frac{e^{2}}{\hbar^{2}} \sum_{{\bf k}} \frac{\partial \varepsilon}{\partial k_{\alpha}} \tau_{\omega} \frac{\partial }{\partial k_{\beta}} f_{0} \left({\bf k} \right).
\end{equation} 
The Pockels coefficient contains a BCD part and a skew-scattering part, $P_{\alpha\beta\gamma} = P_{\alpha\beta\gamma}^{(\mathrm{BCD})} + P_{\alpha\beta\gamma}^{(\mathrm{skew})}$. The BCD Pockels comes from the combination of the anomalous velocity induced by Berry curvature and distribution function with symmetric scattering, ${\bf j}_{\omega, 2, \mathrm{BCD}} (t) =  -  \frac{e^{2}}{\hbar} e^{i \omega t} \sum_{{\bf k}} \boldsymbol{\mathcal{E}}^{(0)} \times {\bf \Omega} f_{0, 1}^{(0)} ({\bf k}, t) - \frac{e^{2}}{\hbar} \sum_{{\bf k}} {\bf E} \times {\bf \Omega} f_{\omega, 1}^{(0)} ({\bf k}, t)$. The coefficient reads
\begin{equation}
P_{\alpha\beta\gamma}^{(\mathrm{BCD})} = \frac{e^{3}}{\hbar^{2}} \left( \tau_{\omega} \varepsilon_{\alpha \gamma \delta} D_{\beta \delta} + \tau \varepsilon_{\alpha \beta \delta}D_{\gamma \delta}  \right), 
\end{equation}
where $D_{\alpha \beta} = \sum_{{\bf k}} f_{0} \left({\bf k} \right) [\partial \Omega_{\beta} /\partial k_{\alpha}]$ is the Berry curvature dipole (BCD)~\cite{Sodemann2015,QiongMa2018,MakKinFai2019}. The other Pockels contribution, the skew-scattering one, comes from the group velocity and the skew-scattering distribution function, ${\bf j}_{\omega, 2, \mathrm{skew}} (t) = - \frac{e}{\hbar} \sum_{{\bf k}}  \nabla_{ {\bf k} } \varepsilon  f_{\omega, 2}^{(1)} ({\bf k}, t)$. The coefficient satisfies
\begin{align}
P_{\alpha\beta\gamma}^{(\mathrm{skew})} &= \frac{e^{3}}{\hbar^{2}} \left\{ \tau_{\omega}^{3} S^{(1)}_{\alpha \gamma \beta } + \tau_{\omega} \tau^{2} S^{(1)}_{\alpha \beta \gamma } + \tau_{\omega}^{2} \left( \tau_{\omega} + \tau \right) S^{(2)}_{\alpha \beta \gamma } \right\} ,  \\
S^{(1)}_{\alpha \beta \gamma } &= - \frac{1}{\hbar} \sum_{{\bf k}} \frac{\partial \epsilon }{\partial k_{\alpha}}  \frac{\partial  }{\partial k_{\beta}} \sum_{{\bf k}^{\prime}} w^{\mathrm{A}}_{{\bf k}{\bf k}^{\prime}} \frac{\partial  }{\partial k^{\prime}_{\gamma}} f_{0} \left({\bf k}^{\prime} \right),  \\
S^{(2)}_{\alpha \beta \gamma } &= - \frac{1}{\hbar} \sum_{{\bf k}} \frac{\partial \epsilon }{\partial k_{\alpha}} \sum_{{\bf k}^{\prime}} w^{\mathrm{A}}_{{\bf k}{\bf k}^{\prime}}  \frac{\partial^{2}  }{\partial k^{\prime}_{\beta} \partial k^{\prime}_{\gamma}} f_{0} \left({\bf k}^{\prime} \right).
\end{align} 
The ``Snap'' contributions to the Kerr effect can be obtained by combining the group velocity with distribution function calculated with symmetric scattering, ${\bf j}_{\omega, 3, \mathrm{Snap}} (t) = - \frac{e}{\hbar} \sum_{{\bf k}} \nabla_{ {\bf k} } \varepsilon f_{\omega, 3}^{(0)} ({\bf k}, t)$,
\begin{align}
\left[ K_{\alpha\beta}^{\gamma\delta} \right]_{\mathrm{Snap}} &= \left(  \tau_{\omega}^{3} +  \tau_{\omega}^{2} \tau +  \tau_{\omega} \tau^{2} \right) J_{\alpha\beta \gamma \delta},  \\
J_{\alpha\beta \gamma \delta} &= - \frac{e^{4}}{\hbar^{4}} \sum_{{\bf k}}  \frac{\partial \epsilon }{\partial k_{\alpha}} \frac{\partial^{3}  }{\partial k_{\beta} \partial k_{\gamma} \partial k_{\delta}} f_{0} \left({\bf k} \right).
\end{align} 
We note that there can an additional contribution to the Kerr effect that originates from Berry connection polarizability corrections to the velocity. In clean metals, Snap Kerr effects can readily dominate over BCP. In our work, we focus on the Snap contributions.

\subsection{The reflection and transmission matrices}
\begin{figure}
\centering
\includegraphics[width=0.2\linewidth]{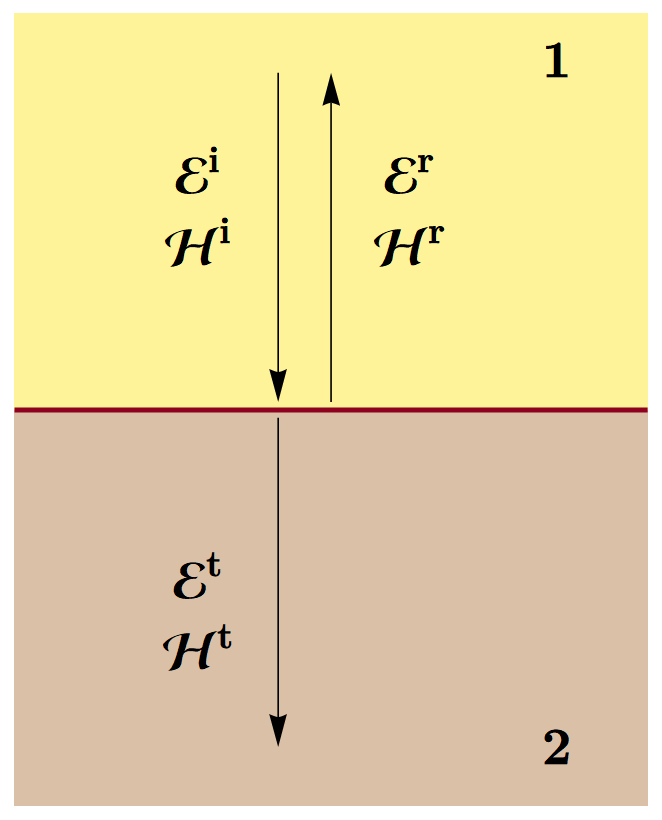}
\caption{Schematic of a 2D thin film (the red line in the middle) with the incident and the reflected electric (magnetic) field $\boldsymbol{\mathcal{E}}^{i}$ ($\boldsymbol{\mathcal{H}}^{i}$) and $\boldsymbol{\mathcal{E}}^{\mathrm{r}}$ ($\boldsymbol{\mathcal{H}}^{\mathrm{r}}$) on side $1$ and the transmitted electric (magnetic) field $\boldsymbol{\mathcal{E}}^{\mathrm{t}}$ ($\boldsymbol{\mathcal{H}}^{\mathrm{t}}$) on side $2$.}
\label{Sfig:Interface}
\end{figure}

Let us consider the electro-optic effect in a 2D thin film modelled as an conducting interface on $x-y$ plane with incident light as plane wave propagating along $z$ direction, i.e., normal incidence. As shown in Fig.~\ref{Sfig:Interface}, we denote the complex amplitude of the incident, the reflected and the transmitted electric field as $\boldsymbol{\mathcal{E}}^{i}$, $\boldsymbol{\mathcal{E}}^{\mathrm{r}}$ and $\boldsymbol{\mathcal{E}}^{\mathrm{t}}$, respectively, and those of the corresponding magnetic field as $\boldsymbol{\mathcal{H}}^{i}$, $\boldsymbol{\mathcal{H}}^{\mathrm{r}}$ and $\boldsymbol{\mathcal{H}}^{\mathrm{t}}$. We number the side with the incident and the reflected as $1$, and the other side with the transmitted as $2$, and we can write the electric field and the magnetic field on the two sides as
\begin{align}
\label{eq:plane-wave-2}
\boldsymbol{\mathcal{E}}^{1} \left(z\right) &= e^{ i \left( k z - \omega t \right) } \boldsymbol{\mathcal{E}}^{i} + e^{ i \left( - k z - \omega t \right) } \boldsymbol{\mathcal{E}}^{\mathrm{r}} = e^{ i \left( k z - \omega t \right) } \left(
\begin{array}{c}
{\mathcal{E}}^{i}_{x} \\
{\mathcal{E}}^{i}_{y}
\end{array}
\right) + e^{ i \left( - k z - \omega t \right) } \left(
\begin{array}{c}
{\mathcal{E}}^{r}_{x} \\
{\mathcal{E}}^{r}_{y}
\end{array}
\right) , \\
\boldsymbol{\mathcal{E}}^{2} \left(z\right) &= e^{ i \left( k z - \omega t \right) } \boldsymbol{\mathcal{E}}^{t} = e^{ i \left( k z - \omega t \right) } \left(
\begin{array}{c}
{\mathcal{E}}^{t}_{x} \\
{\mathcal{E}}^{t}_{y}
\end{array}
\right)  , \\
\boldsymbol{\mathcal{H}}^{1} \left(z\right) &= e^{ i \left( k z - \omega t \right) } \boldsymbol{\mathcal{H}}^{i} + e^{ i \left( - k z - \omega t \right) } \boldsymbol{\mathcal{H}}^{\mathrm{r}} = e^{ i \left( k z - \omega t \right) } \left(
\begin{array}{c}
{\mathcal{H}}^{i}_{x} \\
{\mathcal{H}}^{i}_{y}
\end{array}
\right) + e^{ i \left( - k z - \omega t \right) } \left(
\begin{array}{c}
{\mathcal{H}}^{r}_{x} \\
{\mathcal{H}}^{r}_{y}
\end{array}
\right) , \\
\boldsymbol{\mathcal{H}}^{2} \left(z\right) &= e^{ i \left( k z - \omega t \right) } \boldsymbol{\mathcal{H}}^{t} = e^{ i \left( k z - \omega t \right) } \left(
\begin{array}{c}
{\mathcal{H}}^{t}_{x} \\
{\mathcal{H}}^{t}_{y}
\end{array}
\right),
\end{align}
For simplicity, we assume that the current is only on the 2D interface and that the current-less space is homogeneous. $\boldsymbol{\mathcal{E}}^{i}$, $\boldsymbol{\mathcal{E}}^{\mathrm{r}}$, $\boldsymbol{\mathcal{E}}^{\mathrm{t}}$, $\boldsymbol{\mathcal{H}}^{i}$, $\boldsymbol{\mathcal{H}}^{\mathrm{r}}$ and $\boldsymbol{\mathcal{H}}^{\mathrm{t}}$ are linked by the Maxwell equations in the current-less space, $\nabla \times {\bf E} = - \frac{ \partial {\bf B}}{ \partial t } = - \mu_{0} \frac{ \partial {\bf H}}{ \partial t }$, $\nabla \times {\bf H} = \frac{ \partial {\bf D}}{ \partial t } = \epsilon_{0} \frac{ \partial {\bf E}}{ \partial t }$, where we also assume that the 2D material lies in vacuum. Therefore,
\begin{equation}
\boldsymbol{\mathcal{H}}^{i} = - \frac{ i }{ Z_{0} } \tau_{y} \boldsymbol{\mathcal{E}}^{i} , \quad \boldsymbol{\mathcal{H}}^{r} = \frac{ i }{ Z_{0} } \tau_{y} \boldsymbol{\mathcal{E}}^{r}, \quad \boldsymbol{\mathcal{H}}^{t} = - \frac{ i }{ Z_{0} } \tau_{y} \boldsymbol{\mathcal{E}}^{t} ,
\end{equation}
where $\tau_{y}$ is one of the Pauli matrices. Further, $\boldsymbol{\mathcal{E}}^{1}$, $\boldsymbol{\mathcal{E}}^{2}$, $\boldsymbol{\mathcal{H}}^{1}$ and $\boldsymbol{\mathcal{H}}^{2}$ are constrained by the boundary conditions for electromagnetic field, the tangential component of the electric field being continuous across the interface,
and $\hat{\bf z} \times \left( {\bf H^{ 2 } } \left( 0 \right) - {\bf H^{ 1 } } \left( 0 \right) \right) = {\bf j}$. 

The reflected and the transmitted electric field $\boldsymbol{\mathcal{E}}^{\mathrm{r}}$ and $\boldsymbol{\mathcal{E}}^{\mathrm{t}}$ can be expressed with the reflectance and transmittance tensors, $\bar{r}$ and $\bar{t}$ as $\boldsymbol{\mathcal{E}}^{\mathrm{r}} = \bar{r} \boldsymbol{\mathcal{E}}^{\mathrm{i}}$ and $\boldsymbol{\mathcal{E}}^{\mathrm{t}} = \bar{t} \boldsymbol{\mathcal{E}}^{\mathrm{i}}$, where 
\begin{align}
\label{eq:reflectance-matrix}
\bar{r} = \frac{1}{Y} \left( 
\begin{array}{cc}
-2 Z_{0}^{-1} \sigma_{xx} + \sigma_{xy}^{2} - \sigma_{xx} \sigma_{yy} & -2 Z_{0}^{-1} \sigma_{xy} \\
-2 Z_{0}^{-1} \sigma_{xy} & -2 Z_{0}^{-1} \sigma_{yy} + \sigma_{xy}^{2} - \sigma_{xx} \sigma_{yy} \\
\end{array}
\right),
\end{align}
\begin{align}
\label{eq:transmittance-matrix}
\bar{t} = \frac{2}{Y} \left( 
\begin{array}{cc}
2 Z_{0}^{-2} + Z_{0}^{-1} \sigma_{yy} & -  Z_{0}^{-1} \sigma_{xy}  \\
-  Z_{0}^{-1} \sigma_{xy} & 2 Z_{0}^{-2} + Z_{0}^{-1} \sigma_{xx} \\
\end{array}
\right),
\end{align}
where $Y = \left( 2 \epsilon_{0} c + \sigma_{xx} \right) \left( 2 \epsilon_{0} c + \sigma_{yy} \right) - \sigma_{xy}^{2}$, and $Z_{0} = \left( \epsilon_{0} c \right)^{-1} $ is the vacuum impedance.

\end{document}